\documentclass[preprint]{aastex}
\usepackage{subfigure}
\shorttitle{IFS on B2 0902+34}
\shortauthors{Adams and Hill}
\begin{document}
\title{B2 0902+34: A COLLAPSING PROTOGIANT ELLIPTICAL GALAXY AT z=3.4}
\author{Joshua J. Adams\altaffilmark{1}, Gary J. Hill\altaffilmark{2}, 
and Phillip J. MacQueen\altaffilmark{2}}
\altaffiltext{1}{Astronomy Department, University of Texas at Austin,
Austin, TX 78712; email: jjadams@astro.as.utexas.edu}
\altaffiltext{2}{McDonald Observatory, Austin, TX 78712}
\begin{abstract}
We have used the visible integral-field replicable unit spectrograph prototype (VIRUS-P), a new integral field 
spectrograph, to study the 
spatially and spectrally resolved Lyman-$\alpha$ emission line 
structure in the radio galaxy B2 0902+34 at $z=3.4$. We observe a halo of Lyman-$\alpha$ 
emission with a velocity dispersion of 
$\approx$250 km s$^{-1}$ extending to a radius of 50 kpc.
A second feature is revealed in a spatially resolved region where the line profile shows blueshifted structure. 
This may be viewed as either HI absorption at $\approx$-450 km s$^{-1}$ or secondary emission 
at $\approx$-900 km s$^{-1}$ 
from the primary peak. B2 0902+34 is also the only high redshift radio galaxy with a detection of 
21 cm absorption. Our new data, in combination with the 21 cm absorption, suggest 
two important and unexplained discrepancies. 
First, nowhere in the line profiles of the Lyman-$\alpha$ halo is the 21 cm absorber 
population evident. Second, the 21 cm absorption redshift is higher than 
the Lyman-$\alpha$ emission redshift. In an effort to explain 
these two traits, we have undertaken the first three dimensional 
Monte Carlo simulations of resonant scattering 
in radio galaxies. We have created a simple model with two photoionized 
cones embedded in a halo of neutral hydrogen. Lyman-$\alpha$ photons propagate 
from these cones through the optically thick HI halo until reaching the virial radius. 
Though simple, the model produces the features in the Lyman-$\alpha$ data 
and predicts the 21 cm properties. 
To reach agreement between this model and the data, global infall of the 
HI is strictly necessary. The amount of gas necessary to match the model and 
data is surprisingly high, $\ge 10^{12}M_{\odot}$, an order of magnitude 
larger than the stellar mass. The collapsing structure and large 
gas mass lead us to interpret B2 0902+34 as a protogiant elliptical galaxy. 
This interpretation is a falsifiable alternative to the presence of extended HI shells 
ejected through feedback events such as starburst superwinds. 
An understanding of these gas features and a classification of this system's evolutionary state 
give unique observational evidence to the formation events in massive galaxies.
\end{abstract}



\section{Introduction}
\label{Intro_sec}
\par Progress in understanding high redshift radio galaxies (HzRGs, where we 
arbitrarily consider high to mean z$\gtrapprox$2) will advance our picture 
of both massive galaxy formation and evolution. 
HzRG are thought to be the progenitors of massive local cluster giant elliptical galaxies.
Their radio signal allows detection over a huge range of cosmic time. 
We will only discuss a few 
revelant trends of HzRGs as extensive summaries are given by \citet{McCa93} and \citet{Mile08}. 
Radio galaxies are found to be associated
with some of the largest stellar mass galaxies at all redshifts \citep{Lill84,Pent01}.
The environments of massive radio galaxies
have also been progressively traced to higher redshifts with radio galaxies
typically residing in cluster environments at all redshifts
\citep{Long79,Pres88,Hill91,Cari97,Best00a}.
HzRGs often show extended Lyman-$\alpha$ 
emission up to and beyond their radio emission radii. The Lyman-$\alpha$ 
emission probes warm, ionized gas. Against this emission is often seen line 
profile structure which serves as a probe of the neutral gas through scattering 
and absorption. The properties of Lyman-$\alpha$ blobs (LABs) \citep{Keel99,Stei00,Palu04} and 
the gas halos of quasars \citep{Heck91,Bark03} are morphologically similar to those in HzRGs.
\par Cooling radiation may be an important source of extended Lyman-$\alpha$ radiation, and an indication of 
ongoing galaxy formation as infalling gas continues to build the galaxy's mass \citep{Haim00,Fard01}. 
The most direct signature of a high redshift cooling flow is 
extended Lyman-$\alpha$ emission powered by the decrease in gravitational potential energy from infalling 
material. This is a challenging observation and interpretation as other sources of 
Lyman-$\alpha$ generation are usually stronger 
than the cooling radiation, but \citet{Nils06} and \citet{Smit07} show promising cases. 
Low redshift cooling flows have a long history of investigation primarily through 
X-ray observations \citep[e.g.][]{Fabi94}, but space based observations 
\citep{Kaas01,Pete01,Tamu01} have 
given evidence against most cases of low redshift, strong cooling flows and diminished 
the importance of this mechanism for galaxy growth. 
Most massive galaxy growth is now thought to occur through major 
mergers. \citet{Li07} show by a semi-analytic 
model, for example, the likely growth history of a galaxy like the SDSS quasar J1148+5251 at z=$6.42$ 
\citep{Fan03} where 
gas rich major mergers dominate as the source of the mass growth and a starburst phase 
predates a quasar phase. Observationally checking and challenging theories for the relative contributions 
of mergers and steady gas accretion for different galaxy types and epochs is an important goal. 
\par In terms of evolution, AGN feedback has been proposed \citep{Vala99} as the mechanism to produce 
the tight black hole/stellar mass 
correlation \citep{Mago98,Gebh00,Ferr00} and the difference between the 
theoretical \citep{Kais86} and the observed \citep{Mark98,Arna99} 
relation between cluster temperature and X-ray luminosity. 
Radio galaxies in particular and their powerful jets have been invoked as candidates for 
causing feedback \citep{Inou01,Rawl04,Crot06} by depositing mechanical and thermal 
energy to their surroundings. 
A signature of outflowing gas at a velocity exceeding the halo escape speed would be an 
important observation of high redshift feedback.
\par B2 0902+34 \citep{Lill88} is a powerful high redshift 
($z\approx3.39$) radio galaxy of moderate projected radio diameter (40 kpc, 5$\arcsec$) with a 
huge, diffuse Lyman-$\alpha$ halo 
($\approx100$kpc) and is one of the most thoroughly studied HzRGs. Three observations make B2 0902+34
a unique window into HzRG halos as a class. 
First, 21 cm absorption from HI gas against the radio source continuum has been observed at $z_{21 cm}=3.3962$ 
\citep{Uson91,Brig93,Cody03,Chan04}, but any corresponding feature from this HI population has not shown in the 
Lyman-$\alpha$ emission line structure. This is the only HzRG 
for which 21 cm HI absorption has been repeatably measured with seven other HzRGs giving null detections 
\citep{Roet99,DeBr03}. 
Second, \citet{Cari95} used radio spectral index and polarization
mapping to constrain the approaching, northern radio jet inclination to the observer to between 30$\arcdeg$
and 45$\arcdeg$. This is a particularly low inclination angle for a radio galaxy, but it does offer 
very different lines of sight to the two radio jets and any other structures that may follow 
the radio jets. 
Third, \citet{Vill07} have shown the emission line ratios in B2 0902+34 to be consistent with pure 
AGN photoionization through the 
Lyman-$\alpha$, CIV$\lambda$1549, and HeII$\lambda$1640 lines. A subsample of their HzRGs 
show line ratios that requires stellar photoionization. These systems would have 
peculiar and complicated source distributions for Lyman-$\alpha$ emission, but B2 0902+34 
appears to display the relatively well understood geometry of Lyman-$\alpha$ emission 
powered by an AGN. 
Until recently, longslit spectroscopy had detected only Lyman-$\alpha$, CIV, and HeII emission 
lines \citep{Mart95} and shown no 
sign of a bimodal Lyman-$\alpha$ line profile which is often seen in other HzRGs \citep{vanO97}. Deeper 
spectroscopy and a different slit position angle revealed a spatially resolved 
region with a bimodal Lyman-$\alpha$ line profile \citep{Reul07}, with most of the 
emission still from a simple and quiescent line profile. We will from here forward 
define a quiescent Lyman-$\alpha$ line as one with $\sigma<$1000 km s$^{-1}$ to 
indicate emission which does not display turbulent, radio jet dominated kinematics. \citet{Reul07} 
have also reported a weak [O II]$\lambda$3727 
detection in the galaxy's center. Previous observations in deep Lyman-$\alpha$ narrowband \citep{Reul03} and K broadband 
\citep{Eise92} imaging and spectroscopy of the [OIII] 5007\AA\ line \citep{Eale93} have led some authors to 
tentatively interpret B2 0902+34 as a protogalaxy based on its flat 
spectral energy distribution. \citet{Reul03} showed a diffuse Lyman-$\alpha$ 
morphology with two emission 
peaks roughly perpendicular to the radio axis. Unlike the two other HzRGs in their sample, B2 0902+34 did not 
display filamentary structure in Lyman-$\alpha$.
\par We present the first integral field spectroscopy (IFS) data of Lyman-$\alpha$ emission in B2 0902+34. 
This is the fourth study of HzRGs halos using IFS after the 
works of \citet{Adam97}, \citet{Vill07b}, and \citet{Nesv08}. 
\citet{LeFe96} and 
\citet{vanB05} have used IFS to search for Lyman-$\alpha$ Emitters (LAEs) around HzRGs. With 
knowledge of the full spatial distribution of the bimodal Lyman-$\alpha$ line profile, we create 
a simulation for resonant scattering of the Lyman-$\alpha$ photons through a halo of HI. 
We find a model that reproduces most of the important spatial and spectral properties of our 
IFS observation and strictly requires a collapsing structure with large HI mass. The 
natural explanatory power of this picture for other observations, such as the 21 cm absorption data, leads us 
to claim that B2 0902+34 is a protogalaxy still experiencing initial collapse. 
\par In \S \ref{Obs_sec} we describe our observations. In \S \ref{Mod_sec} we present our new resonant 
scattering model and analyze the impact of our model on 21 cm measurements. 
We show in \S \ref{in_or_out} that our data is best fit through 
resonant scattering against an infalling halo. 
In \S \ref{Conc_sec}, we propose further discriminating observations and review. 
Throughout, we use a standard cosmology with H$_0$=71 km s$^{-1}$ Mpc$^{-1}$, 
$\Omega_M$ = 0.27, and $\Omega_{\Lambda}$ = 0.73. This gives 
look back times of 11.8 Gyr, angular scales of 7.5 kpc arcsec$^{-1}$, 
and a luminosity distance of 30 Gpc for B2 0902+34 at $z=3.4$. 
\section{Observations}
\label{Obs_sec}
\par On 2007 UT March 14 and 15 we observed B2 0902+34 with the visible integral-field 
replicable unit spectrograph prototype (VIRUS-P) \citep{Hill08} on the 2.7m Harlan J. 
Smith Telescope at McDonald 
Observatory. VIRUS-P is a spectrograph with an integral field unit (IFU) of 247 fibers in a hexagonal 
pattern close pack with a 
one third fill factor. 
Fed at f/3.65, the fibers have a diameter of 4.1$\arcsec$ and a 107x107 $\sq\arcsec$ field of view. 
Seeing was $\approx$2$\arcsec$, significantly smaller than the fiber diameter. The first night was nominally 
photometric, while the second was not. 
The instrument configuration gave us 236 working 
fibers spread over the field. Astrometry of the fiber positions has been calibrated to 1.0$\arcsec$ rms
from a fixed, offset guide field.  
Observations of three offset positions (called dithers) fill in the area. We took three exposures of 
1200s at each of six subdither positions. 
A subdither as defined here is a second dithering pattern centered on the fiber intersections of the 
first dither set. We use this 
mode to decrease the astrometric uncertainty of spatially unresolved sources, mitigate local CCD cosmetics, and 
subsample extended emission. 
The wavelength range was 
3400-5680\AA\ with a spectral resolution of 5.0\AA. Wavelength calibration was done with eleven 
arc lamp lines of Hg and Ne yielding a $\sigma_{\lambda} = 0.4$\AA. Twilight flats were used to remove 
pixel-to-pixel 
variation, fiber-to-fiber variation, and the fiber illumination profile. We have used our own custom written 
software 
for all reductions. Notably, this reduction does not perform any resampling or linearization of the data. 
The estimates of the 
sky spectrum were made from neighboring objectless fibers and fit with bsplines \citep{Dier93} in a similar 
manner to 
optimal longslit sky subtraction methods \citep{Kels03}. We estimate a spectrally unresolved 5$\sigma$ line 
flux limit of $7\times10^{-17}$ erg s$^{-1}$ cm$^{-2}$ 
and a surface brightness limit of 24.0 AB mag arcsec$^{-2}$ for wavelengths longer than 4000\AA\ when under 
photometric conditions. 
\par
\begin{figure}
\centering
\epsscale{0.8}
\plotone{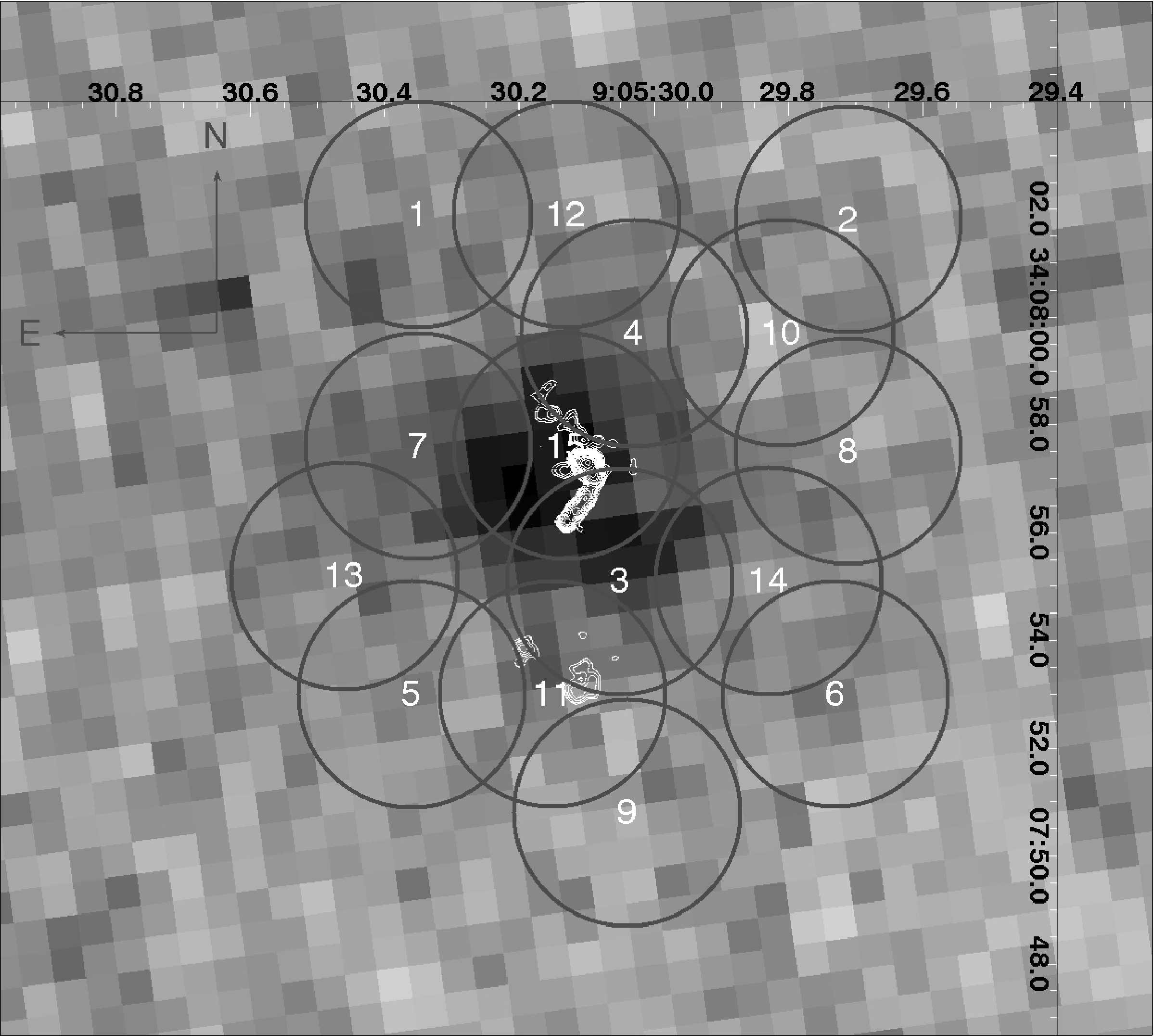}
\caption{The fibers with detected Lyman-$\alpha$ emission are plotted over
a narrowband Lyman-$\alpha$ image of the radio galaxy center. For reference, log scale 
contours are drawn of the 1.65 GHz image at 0.15$\arcsec$ resolution from \citet{Cari95} 
with range 0.36-92.16 mJy/beam. Each fiber has a diameter of 4.2$\arcsec$.}
\label{fig1}
\end{figure}

 We detected Lyman-$\alpha$ emission from the low surface brightness halo of B2 0902+34 in 
fifteen different fiber positions. 
We only include fibers that show emission above a signal-to-noise ratio of 3, calculated from the noise and counts 
within the full width half maximum of a single 
Gaussian fit. We perform Monte Carlo line fitting where 100 realizations of our data sampled from normal distributions 
with our estimated errors produce distributions in our fit parameters from which are extracted central values and dispersions. 
All uncertainty quotes are 1 sigma estimates by this method and similar to the uncertainties from the covariance matrices. 
All quotes of line width 
remove the instrument resolution in quadrature. Table \ref{tab1G} shows 
single Gaussian fits to each fiber's spectrum. Figure \ref{fig1} shows the 
positions of the detections against a narrowband image we have taken and the 
radio data of \citet{Cari95}. 
We confirm the general line profile properties reported in \citet{Reul07} for the region covering their 
slit, but our observations span the entire 2D structure. The entire halo 
contains an emission peak at 5339.0 $\pm$ 2.0\AA\ with a 
FWHM of 600$\pm$90 km s$^{-1}$ which is well fit by a single Gaussian. This primary emission is 
mostly spatially symmetric and smoothly varying in brightness near the 
radio core emission although slightly extended toward the NE. At $\approx15\%$ 
of the strength of the primary emission, 
a secondary emission peak can be fit as a Gaussian at 5324.5$\pm$1.7\AA\ with a FWHM of 630$\pm$270 km s$^{-1}$. 
This secondary emission exists in three fibers on the SW side of the halo. 
Throughout this work, we will refer to any flux 
with a line center above 5330\AA\ as primary emission and 
any flux with a line center below 5330\AA\ as secondary emission. 
For the three fibers with S/N high enough 
for a more complex fit, we give fits in Table \ref{tab2G} where the fitting function consists of two independent 
positive Gaussians. 
Depending on interpretation, this feature may be due to discrete emission sites or scattering and absorption by neutral hydrogen. 
To aid comparison, we give fits in Table \ref{tab1V} with 
a single positive Gaussian and an absorbing Voigt function. Figure \ref{fig2} shows the data and fits to select, 
representative spectra, and Figure \ref{fig3} 
shows the brightness maps of the two spectrally decomposed signatures 
as positive Gaussians. Our data 
show that the longslit spectroscopy of \citet{Lill88}, \citet{Mart95}, and \citet{Reul07} are consistent though only the latter found this secondary, bluer emission 
because of the different slit position angles under their observations and 
the different flux limits. The two dimensional spatial constraints on the secondary emission 
afforded by our data will be crucial to a proper interpretation. We have searched for nearby LAEs in the surrounding fibers but have 
found none within the field. 
We do not make a statistically significant detection of 
continuum in B2 0902+34 which places a lower limit on the 
rest frame equivalent width from the six brightest fibers at 
$>67$ \AA (5 $\sigma$ continuum limit) as shown in Figure \ref{fig_simdet}. 
We have not attempted a flux calibration due to the 
non-photometric conditions, so the aperture to aperture variation in 
surface brightness is overestimated by our data, and the flux limit 
between apertures is not constant. 
\begin{deluxetable}{crrrrrrrr}
\tabletypesize{\scriptsize}
\tablecaption{Individual fiber Lyman-$\alpha$ detections in B2 0902+34\label{tab1G}}
\tablewidth{0pt}
\tablehead{
\colhead{Unique position} & \colhead{RA} & \colhead{Dec} & 
\colhead{Central $\lambda$} & \colhead{Dispersion} 
& \colhead{Normalized} & \colhead{S/N} & \colhead{$\chi^2$} & \colhead{Degrees of}\\ 
\colhead{number} & \colhead{(J2000)} & \colhead{(J2000)} & \colhead{(\AA)} & \colhead{(\AA)} & 
\colhead{Flux} & & & \colhead{Freedom}\\
}
\startdata
1&9:5:30.35&34:8:01.9&5339.24$\pm$7.1&11.5$\pm$9.7&0.27$^{+0.31}_{-0.27}$&3.0&10.8&11\\
2&9:5:29.71&34:8:01.8&5338.96$\pm$0.9&1.1$\pm$1.5&0.19$\pm$0.07&5.2&6.4&3\\
3&9:5:30.05&34:7:55.1&5337.24$\pm$1.0&9.8$\pm$1.1&1.00$\pm$0.11&18.2&25.0&17\\
4&9:5:30.03&34:7:59.7&5338.60$\pm$0.6&3.5$\pm$0.6&0.39$\pm$0.05&9.4&1.1&5\\
5&9:5:30.36&34:7:53.0&5338.84$\pm$2.2&4.4$\pm$3.6&0.40$\pm$0.20&3.3&2.5&4\\
6&9:5:29.73&34:7:53.0&5332.51$\pm$2.8&8.8$\pm$3.2&0.33$\pm$0.11&5.3&13.1&11\\
7&9:5:30.35&34:7:57.6&5338.45$\pm$1.0&7.9$\pm$2.0&0.68$\pm$0.13&10.9&11.3&11\\
8&9:5:29.71&34:7:57.5&5339.27$\pm$1.5&4.3$\pm$2.0&0.26$\pm$0.07&3.9&7.7&5\\
9&9:5:30.04&34:7:50.8&5337.66$\pm$1.3&2.3$\pm$2.2&0.15$\pm$0.11&3.1& 13.9&5\\
10&9:5:29.81&34:7:59.7&5337.92$\pm$1.0&2.5$\pm$1.0&0.20$\pm$0.05&5.1&4.4&4\\
11&9:5:30.15&34:7:53.0&5337.66$\pm$1.3&2.3$\pm$2.2&0.25$\pm$0.09&6.5&5.7&3\\
12&9:5:30.13&34:8:01.9&5336.14$\pm$5.6&6.4$\pm$5.1&0.22$\pm$0.13&4.6&16.0&7\\
13&9:5:30.46&34:7:55.2&5339.12$\pm$2.2&4.0$\pm$1.6&0.19$\pm$0.07&4.3&3.8&3\\
14&9:5:29.83&34:7:55.1&5338.03$\pm$2.9&14.1$\pm$1.9&0.68$\pm$0.16&8.6&39.8&21\\
15&9:5:30.13&34:7:57.6&5338.96$\pm$1.0&4.4$\pm$1.9&0.33$\pm$0.09&7.9&1.7&4\\
\enddata
\end{deluxetable}
\begin{deluxetable}{crrrrrr}
\tabletypesize{\scriptsize}
\tablecaption{Select regions in B2 0902+34 fit with two Gaussians\label{tab2G}}
\tablewidth{0pt}
\tablehead{
\colhead{Unique position} & \colhead{Central $\lambda_1$} & \colhead{Dispersion$_1$} 
& \colhead{Normalized} & \colhead{Central $\lambda_2$} & \colhead{Dispersion$_2$}
& \colhead{Normalized}\\ 
\colhead{number} & \colhead{(\AA)} & \colhead{(\AA)} & \colhead{Flux$_1$}&
\colhead{(\AA)} & \colhead{(\AA)} & \colhead{Flux$_2$}\\
}
\startdata
2&5339.22$\pm$1.5&6.5$\pm$2.5&0.78$\pm$0.27&5323.34$\pm$7.7&3.3$^{+5.0}_{-3.3}$&0.19$^{+0.28}_{-0.19}$\\
5&5338.98$\pm$4.0&4.3$\pm$3.3&0.15$\pm$0.10&5326.97$\pm$3.2&2.8$\pm$2.7&0.14$\pm$0.09\\
12&5344.37$\pm$1.5&4.7$\pm$3.0&0.31$\pm$0.13&5325.49$\pm$1.0&2.9$\pm$1.0&0.22$\pm$0.06\\
\enddata
\end{deluxetable}
\begin{deluxetable}{crrrrrr}
\tabletypesize{\scriptsize}
\tablecaption{Select regions in B2 0902+34 fit with Gaussian and Voigt profile\label{tab1V}}
\tablewidth{0pt}
\tablehead{
\colhead{Unique position} & \colhead{Central $\lambda_e$} & \colhead{Dispersion} 
& \colhead{Normalized} & \colhead{Central $\lambda_a$} & \colhead{b}
& \colhead{Column density}\\ 
\colhead{number} & \colhead{(\AA)} & \colhead{(\AA)} & \colhead{Flux}&
\colhead{(\AA)} & \colhead{(km/s)} & \colhead{(1/cm$^2$)}\\
}
\startdata
2&5335.79$\pm$2.0&8.9$\pm$1.6&1.11$\pm$0.39&5329.86$\pm$1.7&110$^{+275}_{-110}$&8.3$\times{10^{13}}^{+1.4\times10^{14}}_{-8.3\times10^{13}}$\\
12&5335.77$\pm$2.7&8.1$\pm$3.4&1.25$\pm$1.00&5335.40$\pm$1.4&255$^{+262}_{-255}$&2.4$\times{10^{14}}^{+3.8\times10^{14}}_{-2.4\times10^{14}}$\\
\enddata
\end{deluxetable}

\section{Interpretation}
\label{Mod_sec}
\par A bimodal Lyman-$\alpha$ line 
profile may be generated by several effects. First, \citet{Cook08} give imaging and spectroscopy of a Lyman Break Galaxy 
where the line profile is similar to that of B2 0902+34, which they interpret as being due to multiple merging galaxies even 
though the line widths are surprisingly narrow and the morphology undisturbed. 
This possibility in B2 0902+34 cannot be easily discarded. The secondary Lyman-$\alpha$ peak 
and a peak of continuum emission in Figure 8 of \citet{Reul03} appear somewhat coincident. However, our detection of the 
secondary emission peak spans three fibers, which covers a much larger area than the continuum signal. 
We speculate that this continuum may be stimulated or entrained material in the larger radio jet as 
invoked as a leading candidate to explain the alignment effect 
\citep{Bege89,Cham90,McCa93} for radio galaxies. In the case of our supposition, the 
Lyman-$\alpha$ emission in this region originates around the rear radio jet and has a kinematic signature 
unaffected by peculiar motion of the continuum component. The large HI halo and 
scattering process we postulate would also 
cause a frequency redistribution which would wipe out any signature of the original kinematics, 
be they from a merging component or a rear photoionized cone. 
Also, the primary Lyman-$\alpha$ peak does not have a counterpart continuum peak, making it at least unlikely to be affiliated 
with a merger event. 
The second possibility is scattering from discrete regions of neutral hydrogen 
such as an outbound shell due to a starburst superwind, and has been modeled 
and observed for LABs \citep{Ohya03,Wilm05} and Lyman Break Galaxies 
\citep{Shap03}. Third, a robust feature of 
optically thick resonant scattering in static hydrogen clouds 
is a bimodal line profile straddling the rest wavelength \citep{Urba81}. 
The addition of bulk, distributed velocity fields can then process the 
escaping spectra and be used to match a wide range of observed, high redshift 
galaxies \citep{Ahn02b,Verh07,Verh08}. In these models, the source of Lyman-$\alpha$ 
emission is either a central point 
or a spherical distribution surrounded by a scattering medium. None of these pictures can seemingly 
explain the differing line profile seen between B2 0902+34's Lyman-$\alpha$ and 21cm line profiles. 
We submit a fourth mechanism, related to the third mechanism,  
to generate Lyman-$\alpha$ bimodality which is particularly suited to radio galaxies: 
two spatially segregated emission cones 
within a resonantly scattering halo. Resonant scattering effects have been shown through Lyman-$\alpha$ and H$\alpha$ imaging 
to be important in low redshift objects \citep{Haye07} and specifically with bulk velocity structure as the primary 
factor determining Lyman-$\alpha$ escape \citep{Kunt98}. 
\subsection{Physical Assumptions and Constraints}
\par
\begin{figure}
\centering
\epsscale{1.0}
\includegraphics[scale=0.40] {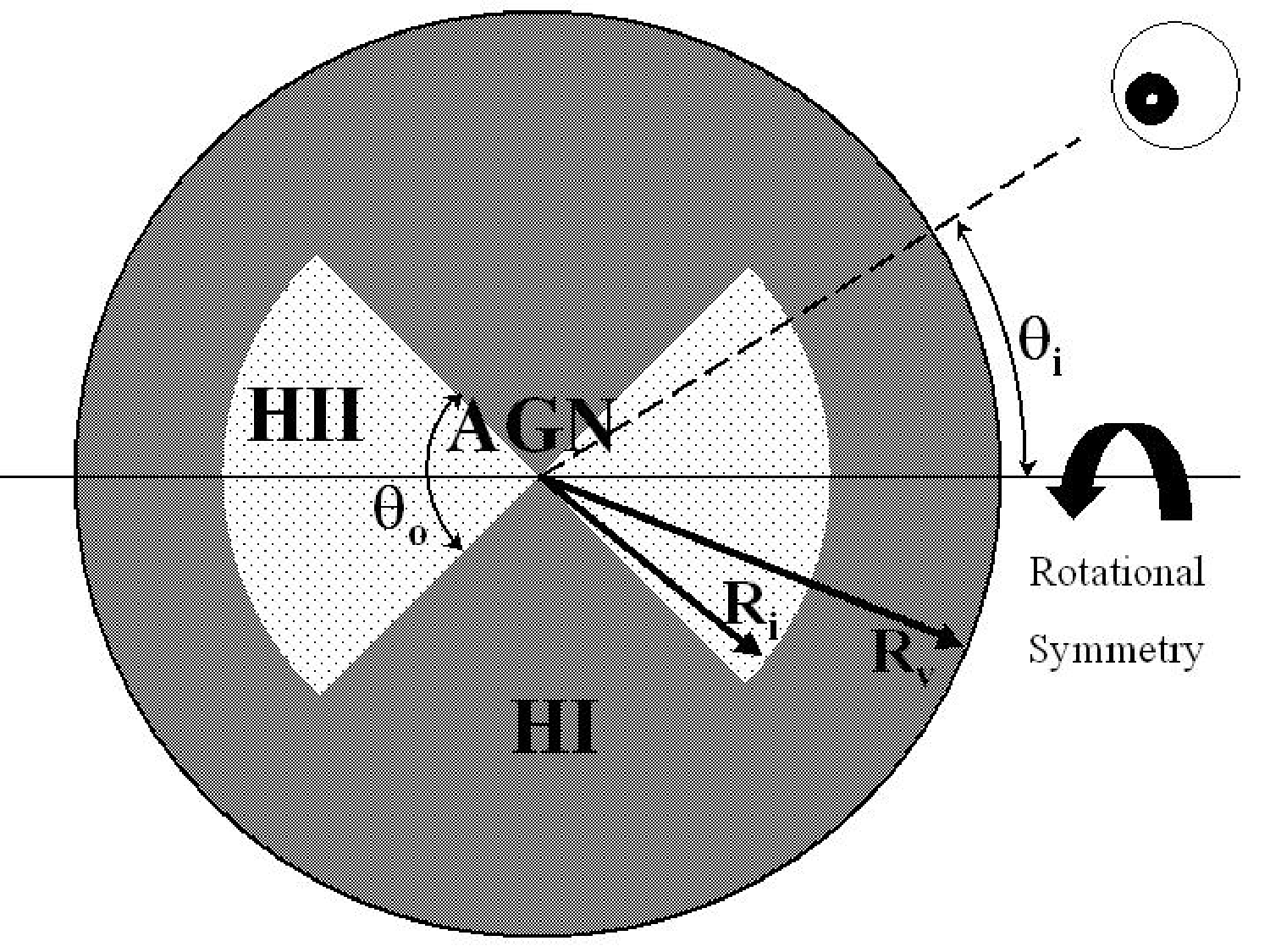}
\caption{The geometry of our simulation. $\theta_i$ is the inclination as constrained by 
\citet{Cari95}. $R_v$ is the system's virial radius. $R_i$ is the ionization radius of the cones. 
$\theta_o$ is the opening angle of the ionization cones assumed here to be $90\arcdeg$. 
$R_i$, $R_v$, $\theta_o$, and two variables controlling the velocity field are the model's five tunable parameters.}
\label{HzRG_cart}
\end{figure}

 The basic geometry we investigate is shown in Figure \ref{HzRG_cart}. This scenario is based on a 
proposal in \citet{vanO97}, although that work does not investigate the key effects of resonant scattering. 
The Lyman-$\alpha$ emission in HzRGs is often strongest 
along the radio axis in a biconal geometry. This geometry has 
been advocated on AGN ionizing photon counting grounds \citep{McCa89} and for AGN 
unification models \citep{Bart89,Anto93}. Surrounding the cones is a large halo of neutral hydrogen gas. 
Our model postulates an opening angle of 90$\arcdeg$ for each cone. Matching the 
spatial location of the bimodal emission in our data and that of \citet{Reul07}, we 
model the projection of the emission cones axis on the sky at 60$\arcdeg$ east of north. It is not clear from the 
radio data exactly where the radio axis projects, but given the large deprojection uncertainty 
in such a low inclination system and 
the complex bending of the radio image, this geometry is reasonable. 
The ionization cone radius could be constrained by the Lyman-$\alpha$ luminosity, 
but since the filling factor is highly uncertain we keep this radius as an open parameter. We model the density 
profile as baryons at 17$\%$ of the total mass following Navarro-Frenk-White (NFW) profiles \citep{Nava97} 
with a thermal core and concentration parameters of c = 5 
following \citet{Dijk06}, hereafter DHS. The input velocity field 
is parameterized as a simple power law with 
$v_{bulk}=v_{amp}\times v_{vir}\times (r/r_{vir})^\alpha$ 
where $\alpha$ is the index, $r$ is radius, $r_{vir}$ is the virial radius, $v_{vir}$ is the virial velocity, and 
$v_{amp}$ is a scaling factor of order one. 
The input parameters and acceptable ranges we explore in common to DHS are total halo mass, $v_{amp}$, 
and $\alpha$. The halo is isothermal at $10^4$K, and the neutral fraction 
in the ionization cones is $10^{-4}$, both appropriate to the common ranges of density in photoionized regions. 
We model a completely smooth distribution of gas as a simplifying assumption. In a 
very similar type of problem, low redshift cooling flows, \citet{Fabi84} have shown how 
smooth flow assumptions adequately model infalling gas. The linear 
relation between the average number of photon scatterings before escape and the 
line center optical depth as shown in Appendix A ensures that gas clumping is irrelevant 
to our results.
\subsection{Resonant Scattering Model}
\par Given our new data of spatially resolved two dimensional Lyman-$\alpha$ emission, we have sufficient 
information to create  
a simple model for this system based on propagation of Lyman-$\alpha$ photons through an optically 
thick halo of HI with geometry as in Figure \ref{HzRG_cart}. 
Lyman-$\alpha$ radiative transfer is a mature field 
\citep{Harr73,Urba81,Neuf91,Loeb99,Ahn00,Ahn01,Ahn02,Zhen02,Rich03,Hans06,Dijk06,
Dijk06b,Verh06,Tasi06,Laur08} 
with both analytic and computational results for simple geometries. 
Broadly, the model will work by emitting Lyman-$\alpha$ photons in two photoionized cones roughly 
aligned with the radio axes. A red and blue emission line relative to systemic will emerge for each cone 
absent a velocity field in the scattering gas giving four 
peaks in the line profile. The far cone's 
emission will be weaker and further from the HzRGs systemic redshift. Global dynamics of infall will 
enhance the two blue emission peaks and suppress the two red emission peaks. Outflow will 
produce a mirror image spectrum about the systemic redshift and suppress the two blue emission peaks and 
enhance the two red emission peaks. The relative wavelengths between the 
stronger, narrower emission from the near cone and the weaker, broader emission from the rear 
cone can then discriminate between infall and outflow even if the 
systemic redshift is unknown. This method provides constraints on the HI mass and velocity field by 
measuring the bimodal line properties and the systemic redshift if available.

\par The study of Lyman-$\alpha$ resonant scattering can be profitably attacked through Monte Carlo methods. 
We base our computations primarily on DHS, where DHS used Monte Carlo methods 
to attempt discrimination between energy sources and conditions in the data on LABs. Our version 
of the Monte Carlo resonant scattering code differs in three ways from their description. 
First, we do not include deuterium in our simulations. Although 
they show convincingly that deuterium is important for certain wavelengths and conditions, our very high optical 
depth and temperature environments make its contribution negligible and a needless drain on computing performance. 
Second, all of DHS's geometries are restricted to spherical symmetry. As we will invoke the two 
cones of the AGN as coaxial with the cones of Lyman-$\alpha$ emission, our models are without spherical symmetry. 
Our simulations therefore are a function of inclination angle and require the same number of simulation 
photons in each inclination angle bin as in an entire DHS simulation to reach an equal S/N. Third, 
DHS have a geometry that yields an analytic expression between optical depth and traversed physical distance 
for thin shells of constant density and velocity gas. 
The general density and velocity field conditions in our model mean that we must integrate this distance numerically. 
We use a root finding algorithm with a relative accuracy of $10^{-8}$ on a fifth order Runge-Kutta code to calculate 
optical depths and sites of scatter. 
It is this generalization that enables us to pursue arbitrary density, temperature, and 
velocity fields. It will be possible to leverage this generalization by coupling our code to 
the more realistic conditions output by hydrodynamic codes in future work, but for now we use 
relatively ad hoc parameterizations. 
We used, in common with DHS, an acceleration scheme with x$_{crit}=3$ as described in their 
\S3.1.1. 
Our simulation work is an extension and application of the 
powerful method of DHS to a different class of object, HzRGs. 
We have run our code for the four test cases shown in DHS's Figure 1 and found agreement 
as demonstrated in Appendix A. 
\par Our main differences in operation from the DHS models are the changes in HI density in the presence 
of ionization cones and AGN photoionization as the dominant Lyman-$\alpha$ energy source 
determining the emissivity profile. Lyman-$\alpha$ emission powered by a cooling flow and its associated 
uncertainties are unnecessary in B2 0902+34 with the strong evidence for AGN photoionization. 
Within the ionization cones, Lyman-$\alpha$ is generated randomly in position following 
the recombination rate. 
\par Our simulation ends when a photon leaves the virial radius. Further radiative transfer effects in the IGM 
are beyond this work's scope, although they can be important \citep{Dijk06b,Bark04}. We briefly 
discuss the qualitative impact that IGM processing by infalling neutral hydrogen outside the virial radius 
could have to our model. Figure 2 of \citet{Dijk06b} 
gives transmission curves at $z=3$ for a variety of impact parameters and for cases with and without 
a nearby bright, ionizing quasar. Multiplying this transmission into a single peaked emission profile 
can produce a bimodal line profile, but this is not a likely explanation by itself for B2 0902+34. 
Only emission lines that 
are well centered around the systemic redshift can be modified by the IGM into a bimodal profile as 
the transmission minimum occurs slightly redward of systemic. All our models produce Lyman-$\alpha$ emission 
that is blueshifted enough that the IGM processing could not create a bimodal spectrum. The higher 
21 cm redshifts also imply that the Lyman-$\alpha$ emission is not near the systemic redshift. 
We also do not observe the spatial dependence with impact parameter 
that IGM processing would create. For these reasons and scope, we neglect IGM processing. 
\par We must review a few of the broad findings in DHS to understand our models. Radiative escape from a static 
gas cloud will produce a bimodal spectrum. A photon has a much longer mean free path when it is at an off-resonance wavelength, 
which the thermal kicks in a scatter can cause. This situation will produce peaks redward and blueward 
of systemic with higher optical depths pushing the peaks out to 
more separated wavelengths. If the same situation has global infall (outflow) in the scattering gas, 
the photon moving outward and being 
scattered will have on average a blue-(red-)shift in the frame of the atom. A photon formerly in the 
red (blue) wing will move into resonance and have its travelling distance supressed, while a photon formerly 
in the blue (red) wing will be pushed further into the scattering wing and travel further before its next scatter. 
In practice, the suppressed emission peak is likely to be far below detection limits for the range of optical depths 
we investigate. Therefore we return to a solitary emission line with no information remaining on the system's 
systemic redshift for cases of spherical symmetry. As DHS acknowledge, there is a degeneracy in observation between systems in infall 
and outflow. We turn to a stark physical difference between the DHS modeled LABs and our HzRGs to solve this problem. HzRGs frequently 
have two discrete regions of emission as ionization cones powered by the AGN. In a low inclination angle system such as B2 0902+34, 
the optical depth difference between these two regions to an observer can be substantial. Such a system should exhibit 
a weaker, spatially restricted emission profile caused by the rear ionization cone superimposed on a 
stronger, nearly circularly distributed emission profile caused by the near ionization cone. 
The relative order in wavelength between the weak and strong emission peaks will point in the direction of the 
unobserved, systemic redshift. So, HzRGs may be unique as high redshift halos where 
observation can distinguish without ambiguity between infall and outflow, solely from Lyman-$\alpha$ 
emission profiles. B2 0902+34 provides a further test of this idea since it is the 
only HzRG with a detection of 21 cm HI absorption.
\subsection{Evaluation of Model}
\label{mod_eval}
\par
\begin{figure}
\centering
\epsscale{1.0}
\subfigure{\includegraphics[angle=270,scale=0.3] {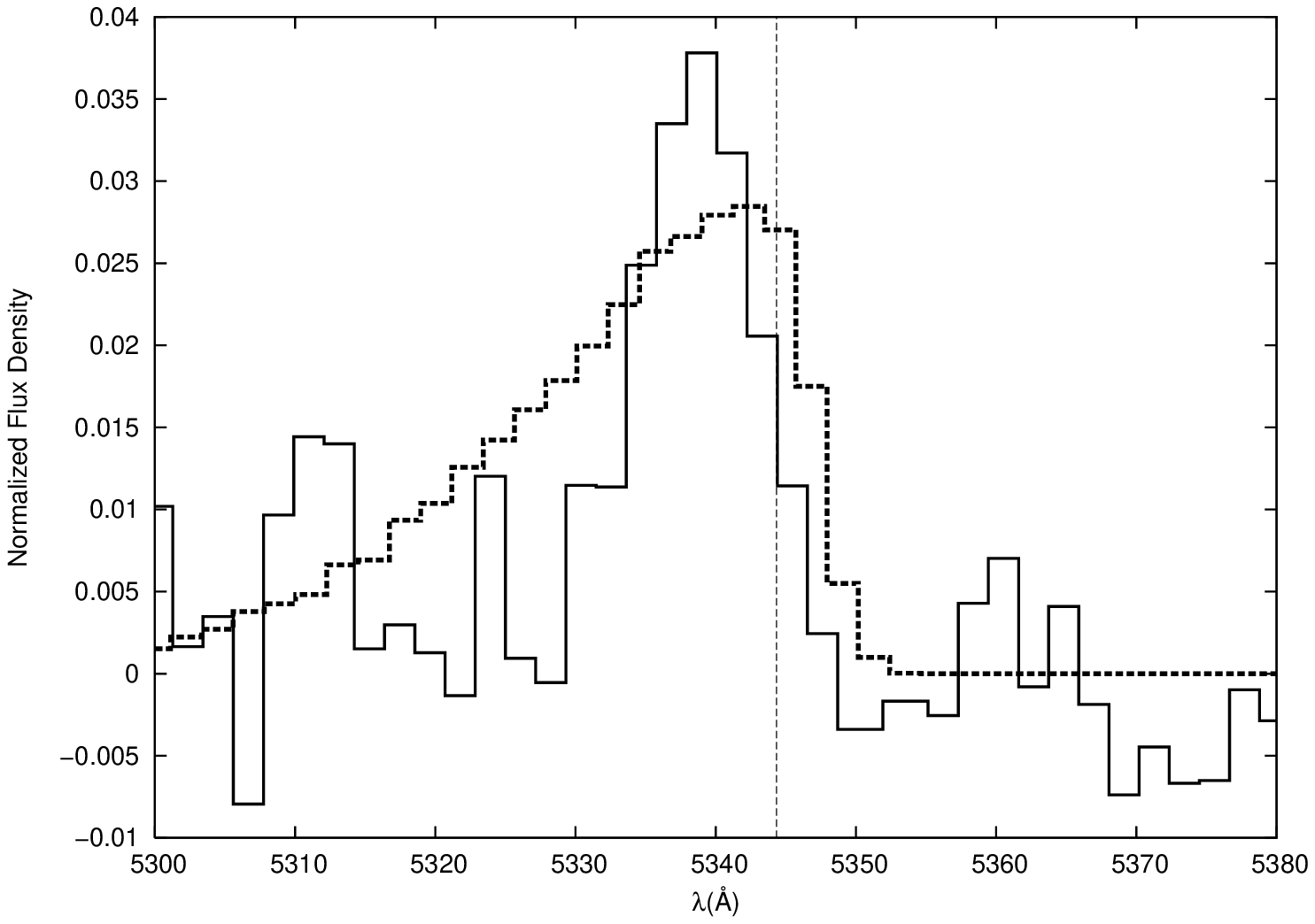}}
\\
\subfigure{\includegraphics[angle=270,scale=0.3] {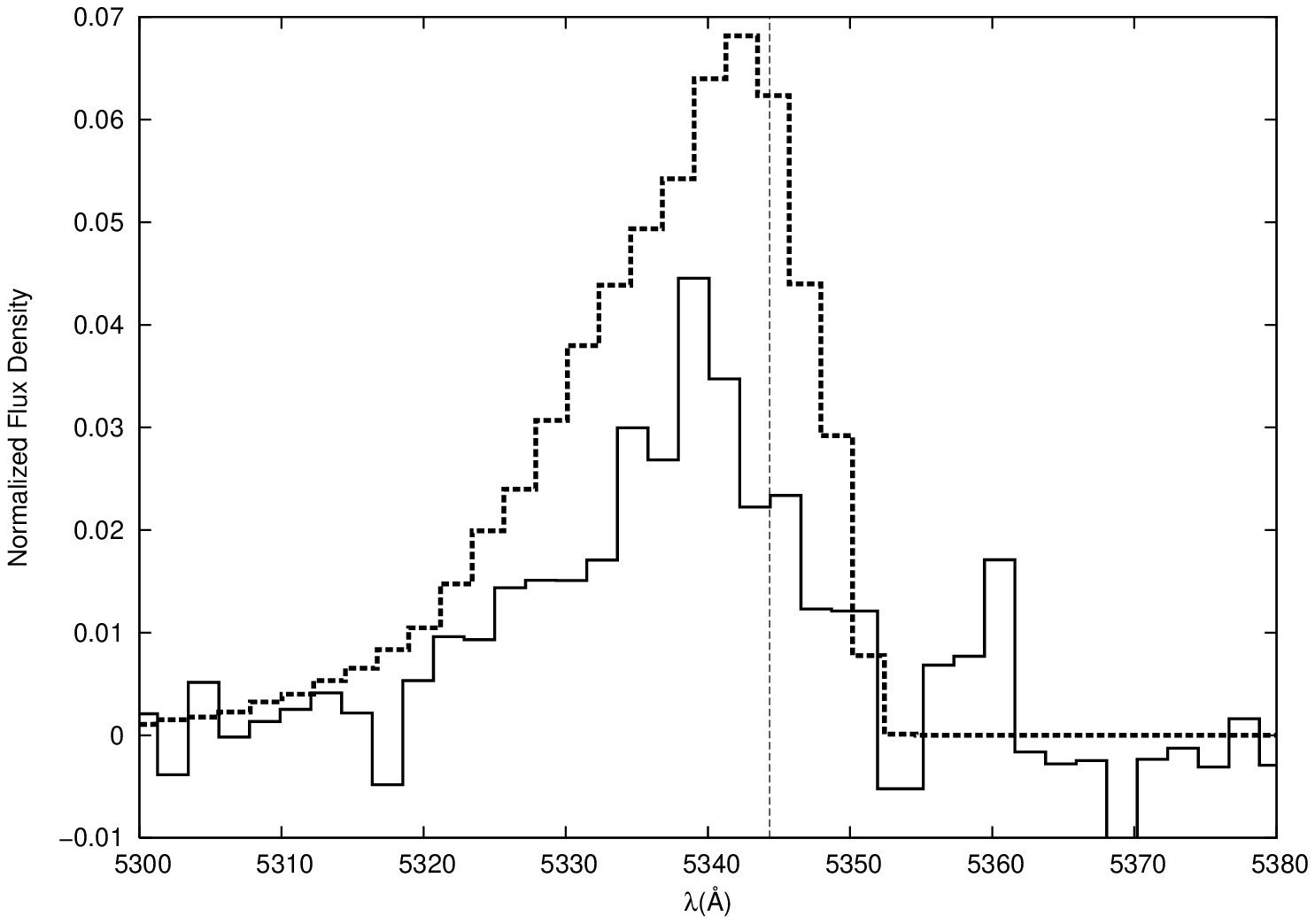}}
\\
\subfigure{\includegraphics[angle=270,scale=0.3] {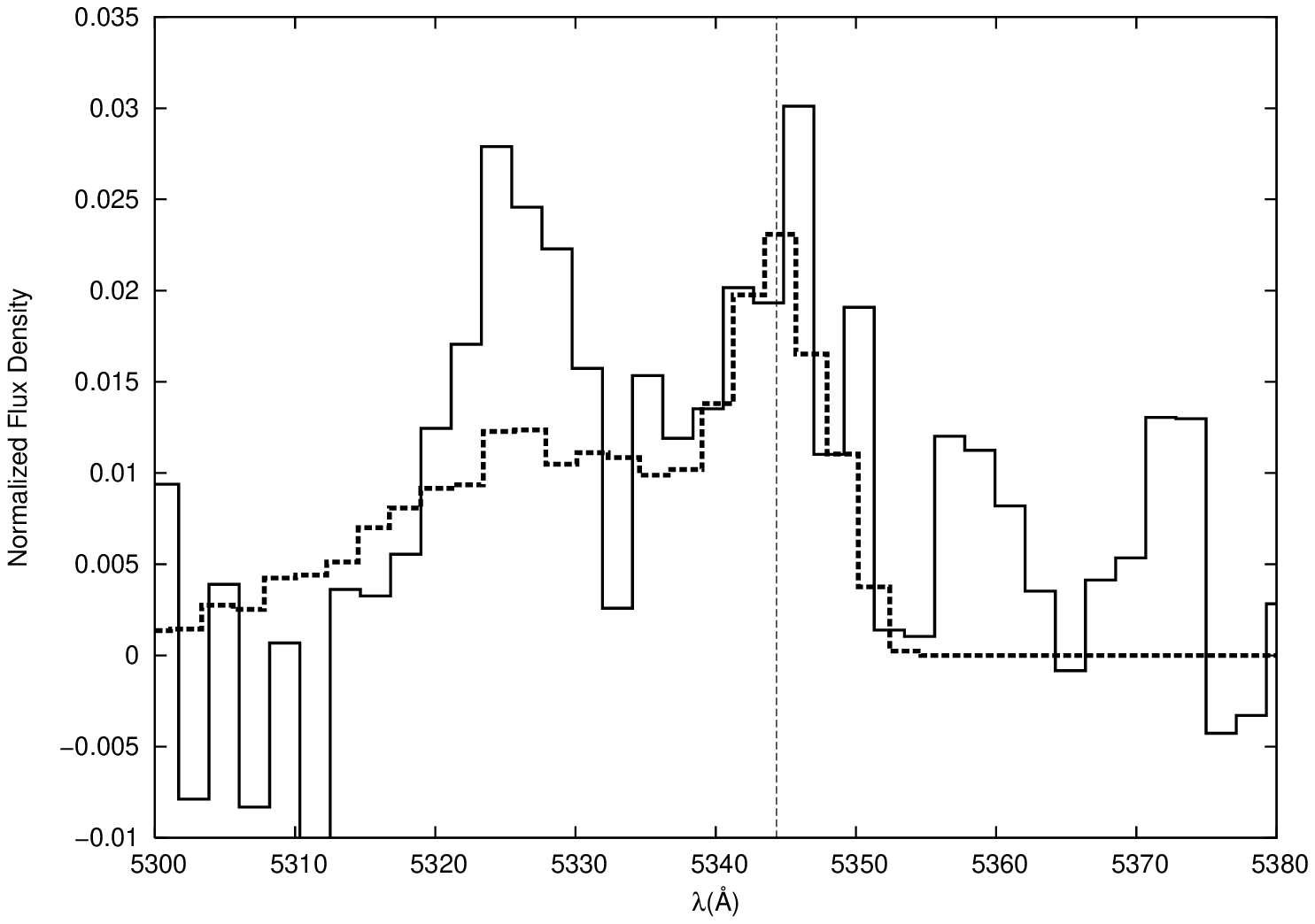}}
\subfigure{\includegraphics[angle=270,scale=0.3] {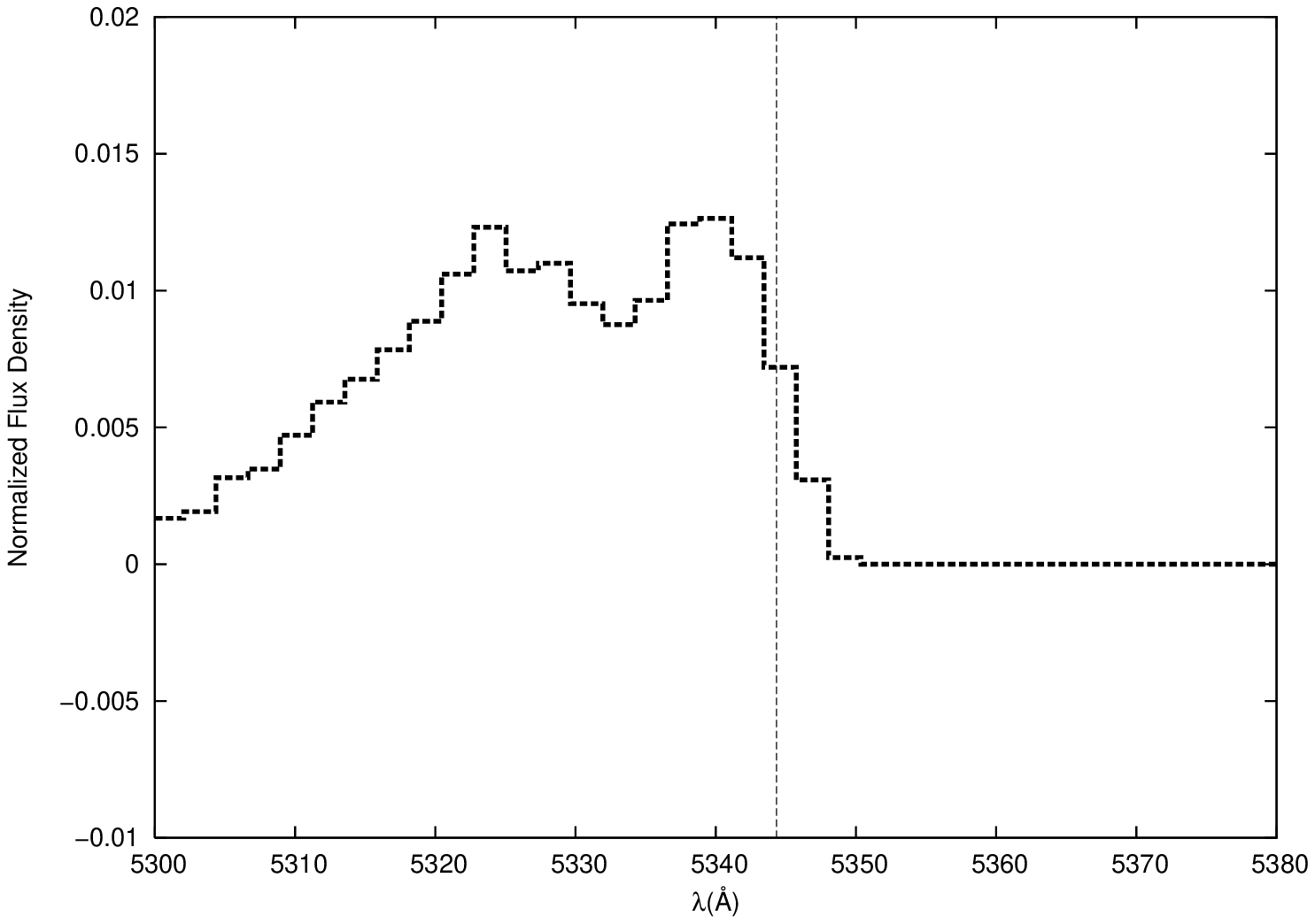}}
\caption{The spectra of both our IFS data and resonant scattering simulation 
from select fibers. 
The thin dotted line indicates the redshift measured for 21 cm absorption.
\textit{Solid} Observed. \textit{Thick dotted} 
Simulated. \textit{Top} Fiber 4, representative of most of the halo 
area where neither cone directly projects. \textit{Middle} Fiber 7, representative of 
regions spanned by the near photoionized cone. \textit{Bottom left} Fiber 14, representative of 
regions spanned by the rear photoionized cone. Only these regions show a bimodal line profile. 
\textit{Bottom right} A region in the simulation near fiber 14. While we did not have 
a fiber on this exact position, the bimodal spectrum here is a good match to its 
neighbor in fiber 14.}
\label{fig2}
\end{figure}

 In Figure \ref{fig2} we give the observed and simulated spectra in identical apertures for a particular model. 
There are five tunable parameters to our model: the virial 
radius, the ionization radius, the velocity strength, the power law index to the velocity field, and the photoionized 
cones' opening angle. We have coarsely 
run models over the likely halo parameter space, but we only present one model which captures many of the necessary 
observed traits. 
This model realization uses 700,000 photons generated in a $6\times10^{12}$M$_{\sun}$ halo which corresponds to a 
virial radius of 134 kpc and a virial velocity of 437 km s$^{-1}$. The ionization radius extends 
to 101 kpc for this realization. 
We use an infall model where the velocity varies as the radius to the power $\alpha$ where 
here $\alpha=0.5$ and with a 
maximum radially inward velocity at the virial radius of 656 km s$^{-1}$. The cone opening angle 
is 90$\arcdeg$.  
We have one further parameter which we hold constant as it is well constrained by a previously discussed observation: 
the system's inclination angle. We believe $\theta_{i}=45\arcdeg$ as most likely given the 
absence of an observed broad line region, so in practice we bin 
the spatial projections and wavelengths for all photons that exit within 35$\arcdeg$ to 
55$\arcdeg$ of an ionization cone axis. For our model, we have assumed a systemic redshift of 
$z_{systemic}=3.3990$. We have 
done this after judging the velocity offset that the 21 cm absorption measurement should have in \S \ref{sub_21}. 
The various redshifts are reviewed in Table \ref{tabredsh}. 
\par This model reproduces many features of the observation. Most importantly, the 
size of the split in wavelength between the primary and secondary peak is recovered in the 
bottom panel of Figure \ref{fig2}. 
The relative 
intensities between the primary (near cone) and secondary (rear cone) emission are also fairly well matched. 
The regions of the halo not in projection of the receding radio cone also show the proper behavior as demonstrated 
in the top plot of Figure \ref{fig2}. The emission there is not bimodal, and the peak is nearly constant at the 
same wavelength as the redder peak in the bimodal region. Several features in the model clearly do not meet the 
data, and our model is not to be considered the optimum match in parameter space but instead an early attempt. 
The model data all display a blueward skew, especially in Fiber 4 (top panel, Figure \ref{fig2}) 
and regions away from either photoionized cone. 
Our data's S/N does not allow a useful measure of skew, but we predict that 
a much deeper and perhaps higher spectral resolution observation would show skew. This lack of skew is 
currently the most outstanding mismatch between our model and data. 
\par
\begin{figure}
\centering
\epsscale{1.0}
\mbox{
\subfigure[Observed]{\includegraphics[angle=0,scale=0.28] {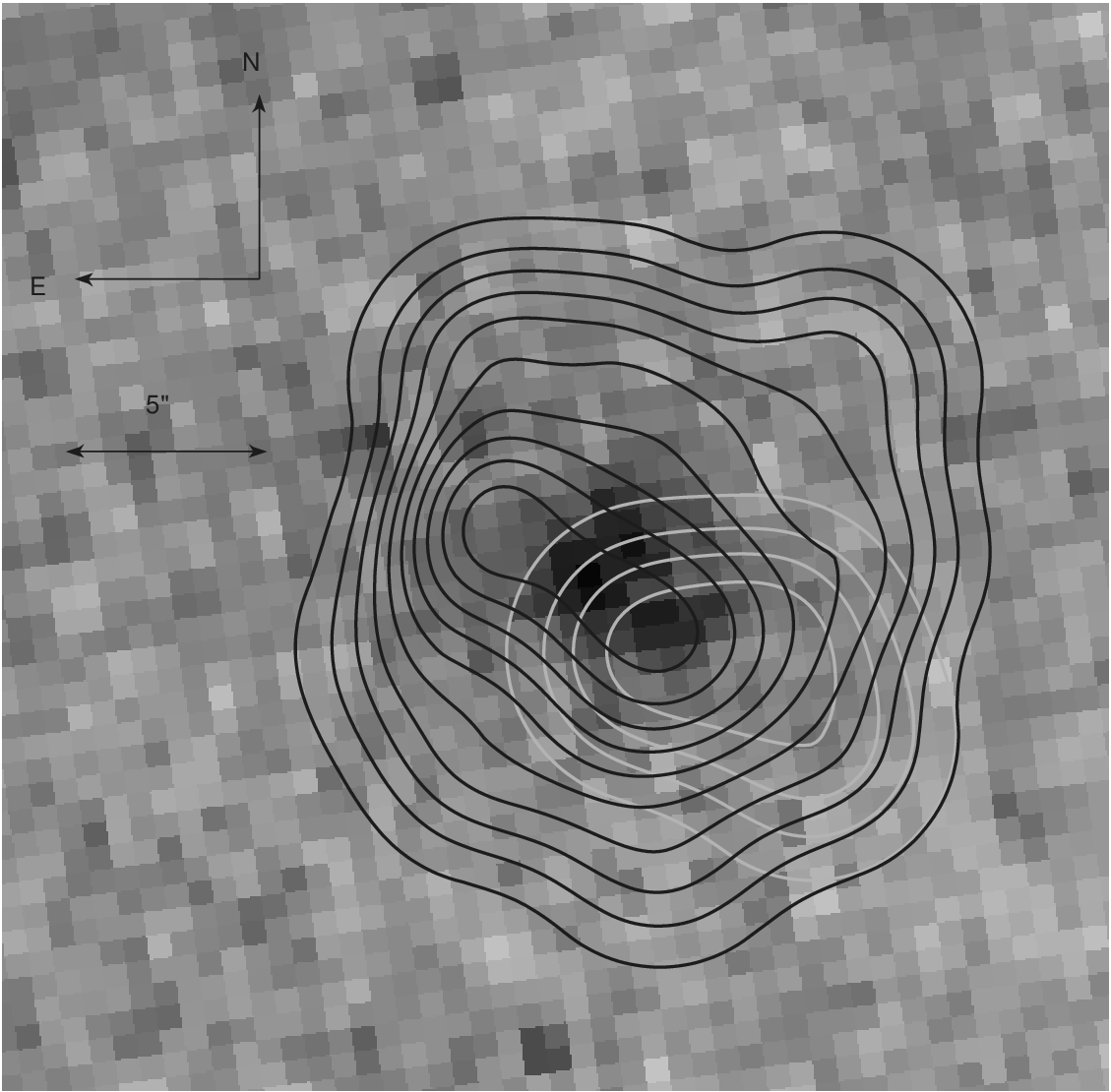}}
\subfigure[Simulated, Spectral Decomposition]{\includegraphics[angle=0,scale=0.28] {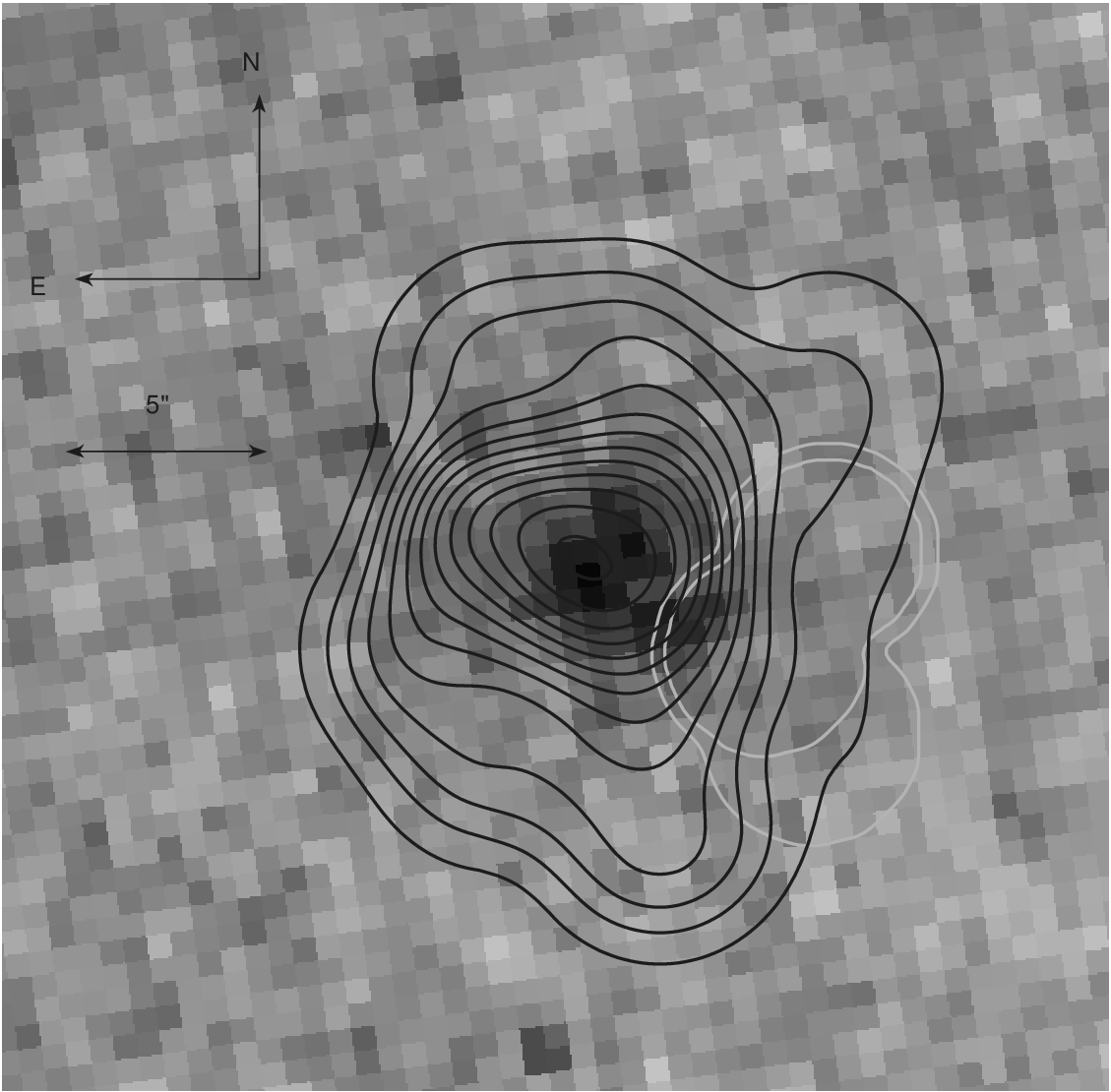}}
\subfigure[Simulated, Origin Cone Decomposition]{\includegraphics[angle=0,scale=0.28] {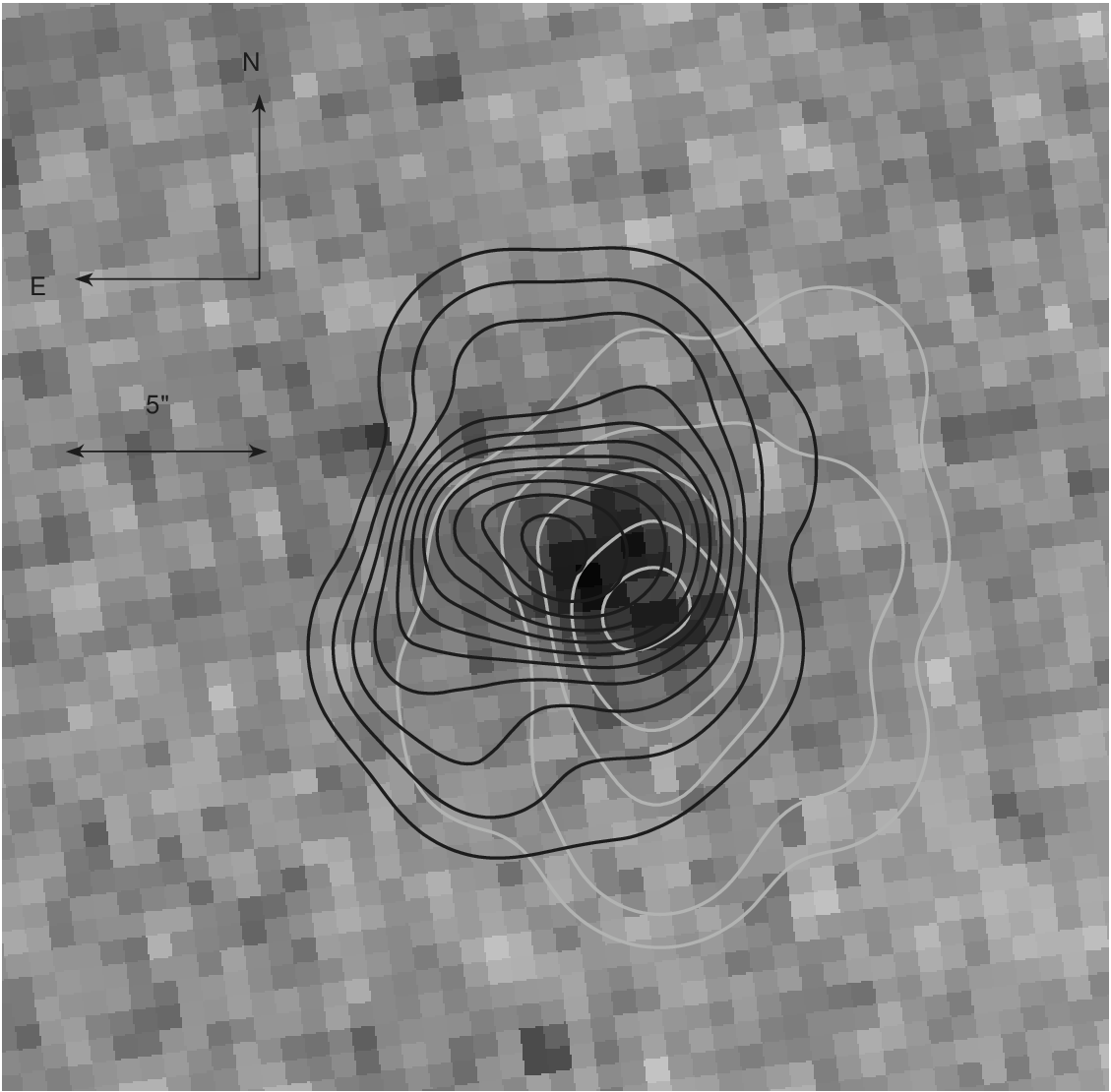}}
}
\caption{Log contours of surface brightness. \textit{Left} Fits to the data. 
The primary and secondary peaks are respectively dark and light. 
\textit{Middle} The simulations described in \S \ref{Mod_sec}, where the same type of 
spectral decomposition as for the data between primary and secondary is performed. \textit{Right} 
The same simulations, but now with a decomposition regardless of the spectral signature but 
where the dark contour follows emission from the near photoionized cone and the light contour 
from the rear photoionized cone. This optimistically represents a clean detangling of spectral 
signal that a model with more optimized parameters may accomplish.}
\label{fig3}
\end{figure}

\begin{figure}
\centering
\epsscale{0.6}
\plotone{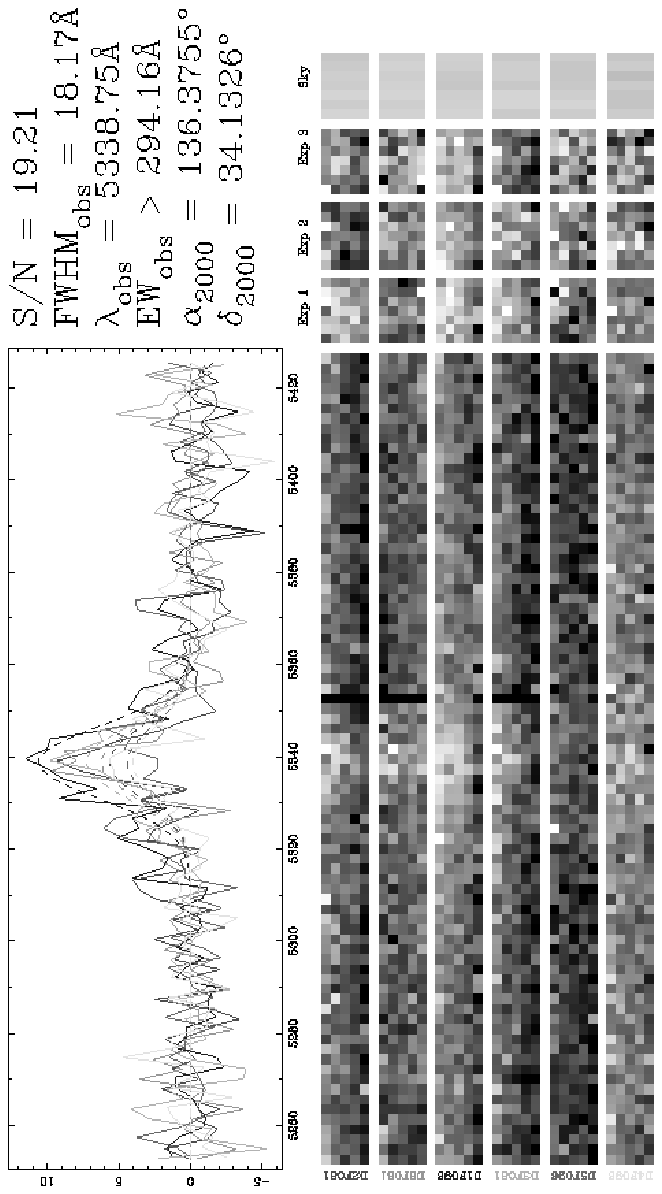}
\caption{The joint detection of Lyman-$\alpha$ emission in a 
subset of the central fibers. The equivalent width limit 
is shown. Each row in the figure's bottom half corresponds to the 
spectrum from a fiber with a unique position on sky. In the labeling of 
Figure \ref{fig1}, these fibers are numbers 4, 15, 3, 7, 14, and 11 from 
top to bottom.}
\label{fig_simdet}
\end{figure}

 We consider our results as surface brightness distributions in Figure \ref{fig3}. The left panel shows the 
observed decomposition between the primary and secondary peaks. The middle panel shows a similar 
decomposition for our model, where emission is decomposed spectrally. 
The right panel shows the model with the emission decomposed by the photoionized cone of origin. 
When we fairly decompose the model as in the middle panel, the region of secondary emission 
is pushed out slightly from the HzRG center and the relative intensity from secondary emission 
is underpredicted. Other models we have run have matched the spatial contours more closely, but have other 
failings such as underpredicting the bimodal wavelength separation, giving much broader lines, 
or showing even more skew. The modest S/N's so far observed for this dim object will 
inevitably cause confusion between skew, line width, and a bimodal line profile, so our 
unoptimized model parameter choices are appropriate to the state of certainty in the current data. 
The spatial distributions and relative intensities 
are fairly well matched at the level of the coarse spatial sampling allowed by our observation's fiber size. The 
offset spatial peak from the near ionized cone is also a robust match and explains why the 
Lyman-$\alpha$ peak and the radio core do not coincide.
\par Other parameter choices have some important trends. We cannot use a much smaller total halo mass, 
because the rest frame wavelength separation in the bimodal regions drops from 4\AA\ to 1\AA\ for a total 
halo mass of $3\times10^{12}$M$_{\odot}$. Larger halo masses are allowed and give more freedom 
in the other model parameters to reach the crucial rest frame wavelength separation of 4\AA, so 
we consider our gas mass to be a lower limit. 
Larger $\alpha$ values produce narrower emission lines with 
more skew, while a value of -0.5 creates lines of huge (40\AA) dispersion and spectral shifts 
(thousands of km s$^{-1}$ from systemic) that are incompatible with the data. Smaller $R_i$ to $R_v$ 
ratios seem to also lower the rest frame 4\AA\ line profile separation unacceptably. A smaller value of 
$\theta_o=60\arcdeg$ gives the appealing feature of more cleanly separating the near and rear 
photoionized cones spectrally much as the right panel of Figure \ref{fig3} shows, but the 
emission not in projection of the near cone begins to show line centers improperly 
centered around the wavelength of secondary emission contrary to observation. The most robust 
result of these models is that a rest frame line separation of 4\AA\ can be accomplished with 
resonant scattering from HzRG cones with a very large HI mass ($10^{12}$M$_{\odot}$), but 
a separation of much more (say, 8\AA) can not without 
masses an order of magnitude larger.  
\begin{deluxetable}{crrrrrr}
\tabletypesize{\scriptsize}
\tablecaption{Redshift estimates in B2 0902+34\label{tabredsh}}
\tablewidth{0pt}
\tablehead{
\colhead{Primary Lyman-$\alpha$} & \colhead{Secondary Lyman-$\alpha$} & \colhead{21 cm absorption} 
& \colhead{Systemic from model}\\ 
}
\startdata
3.3918&3.3795&3.3962&3.3990\\
\enddata
\end{deluxetable}

\subsection{21 cm Predictions}
\label{sub_21}
\par Any correct physical model of B2 0902+34 must take account of the 21 cm absorption data. 
Table \ref{tab1V} shows that fits of absorption to the Lyman-$\alpha$ line profile do not match 
the absorbing population of HI evident in the 21 cm data. 
\Citet{Cody03} detect an absorption feature of FWHM 120 km s$^{-1}$ at z = 3.3962 with 
a column density of $3\times10^{21}$ cm$^{-2}$ under an assumed spin temperature of 
T$_s$=10$^3$K. 
The relative velocities between the Lyman-$\alpha$ emission and the 21 cm absorbing HI and the absence 
of a signature from the 21 cm absorping gas in the Lyman-$\alpha$ profile both give 
strong constrains on possible geometries and conditions. 
\par
\begin{figure}
\centering
\epsscale{1.0}
\subfigure{\includegraphics[angle=0,scale=0.4] {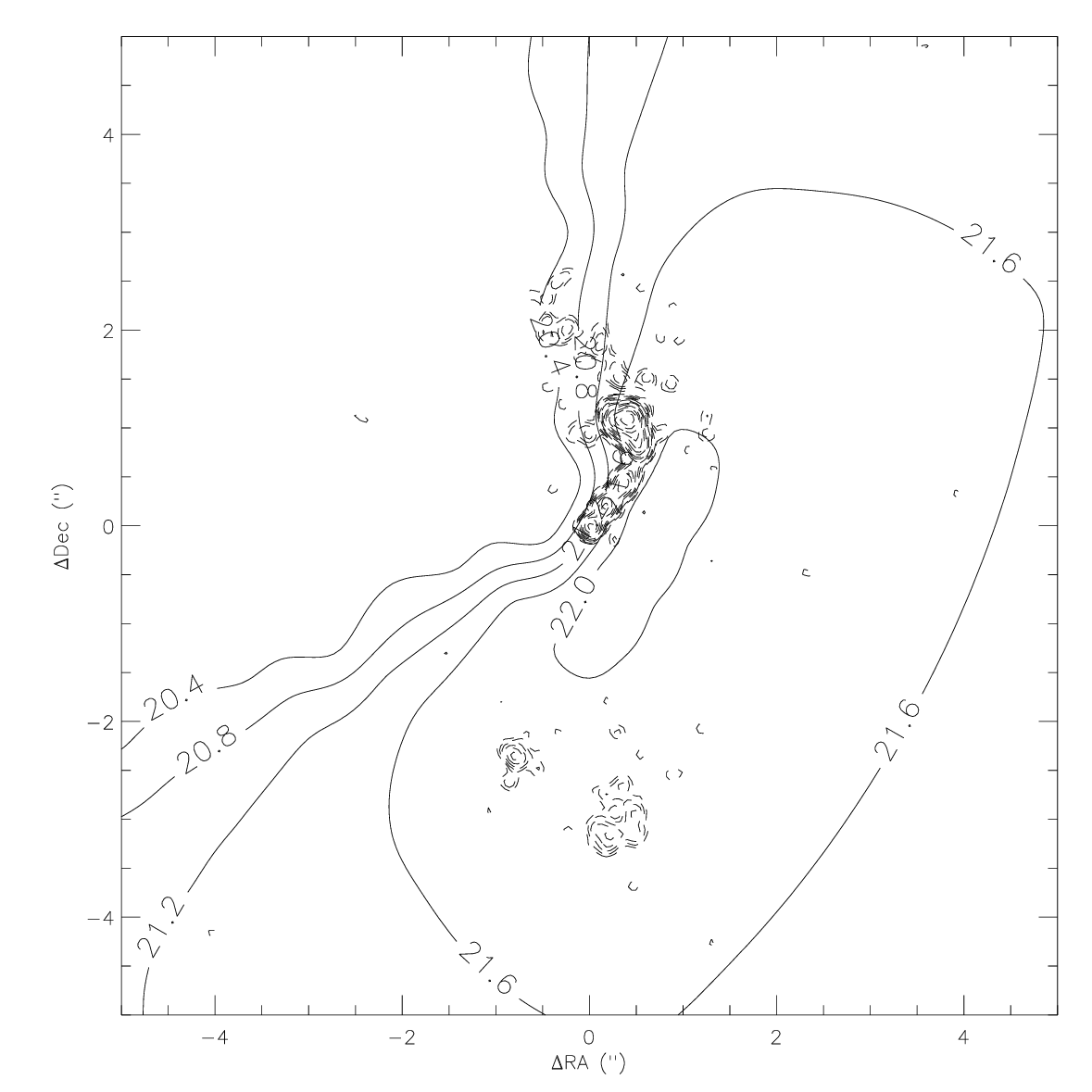}}
\subfigure{\includegraphics[angle=0,scale=0.4] {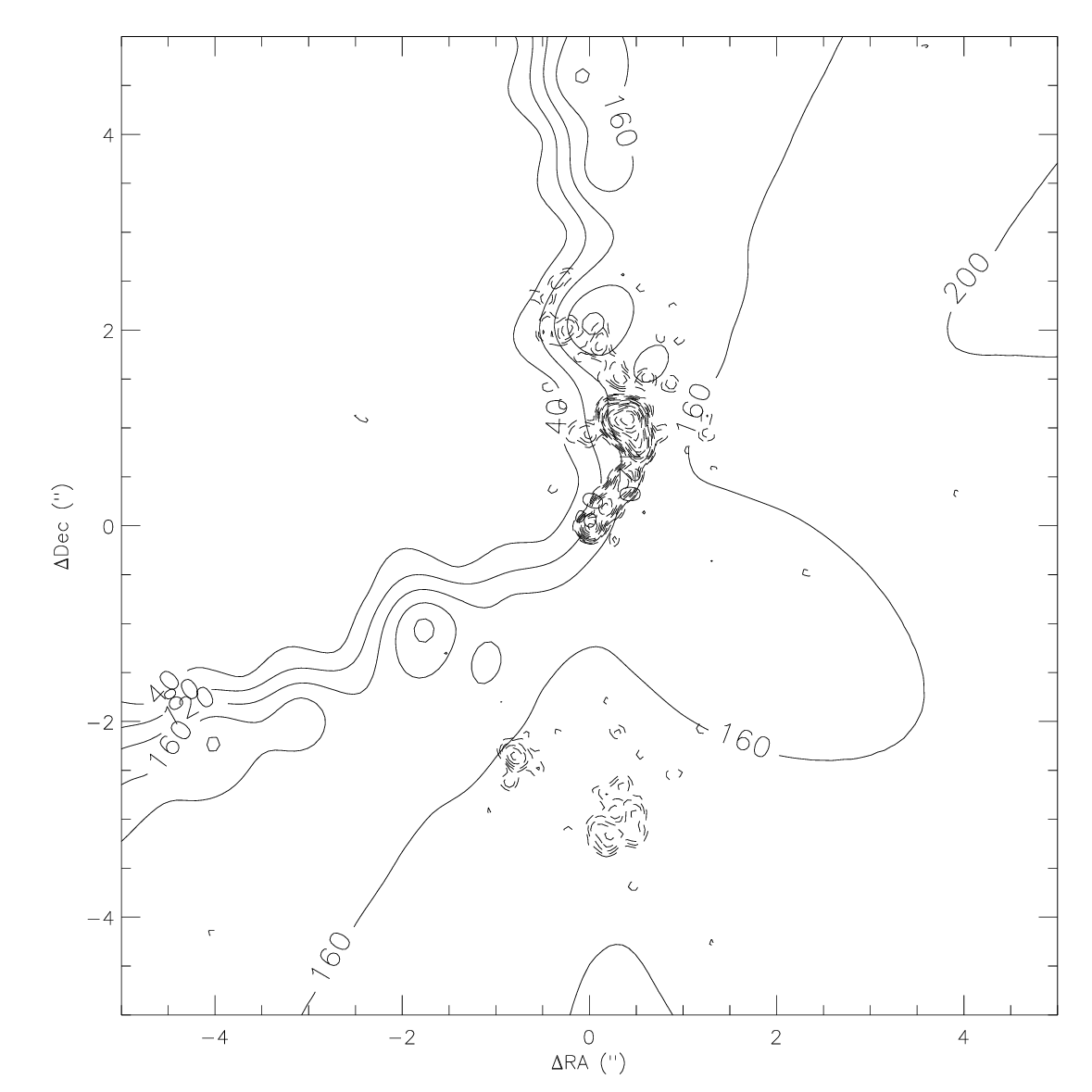}}
\\
\subfigure{\includegraphics[angle=0,scale=0.4] {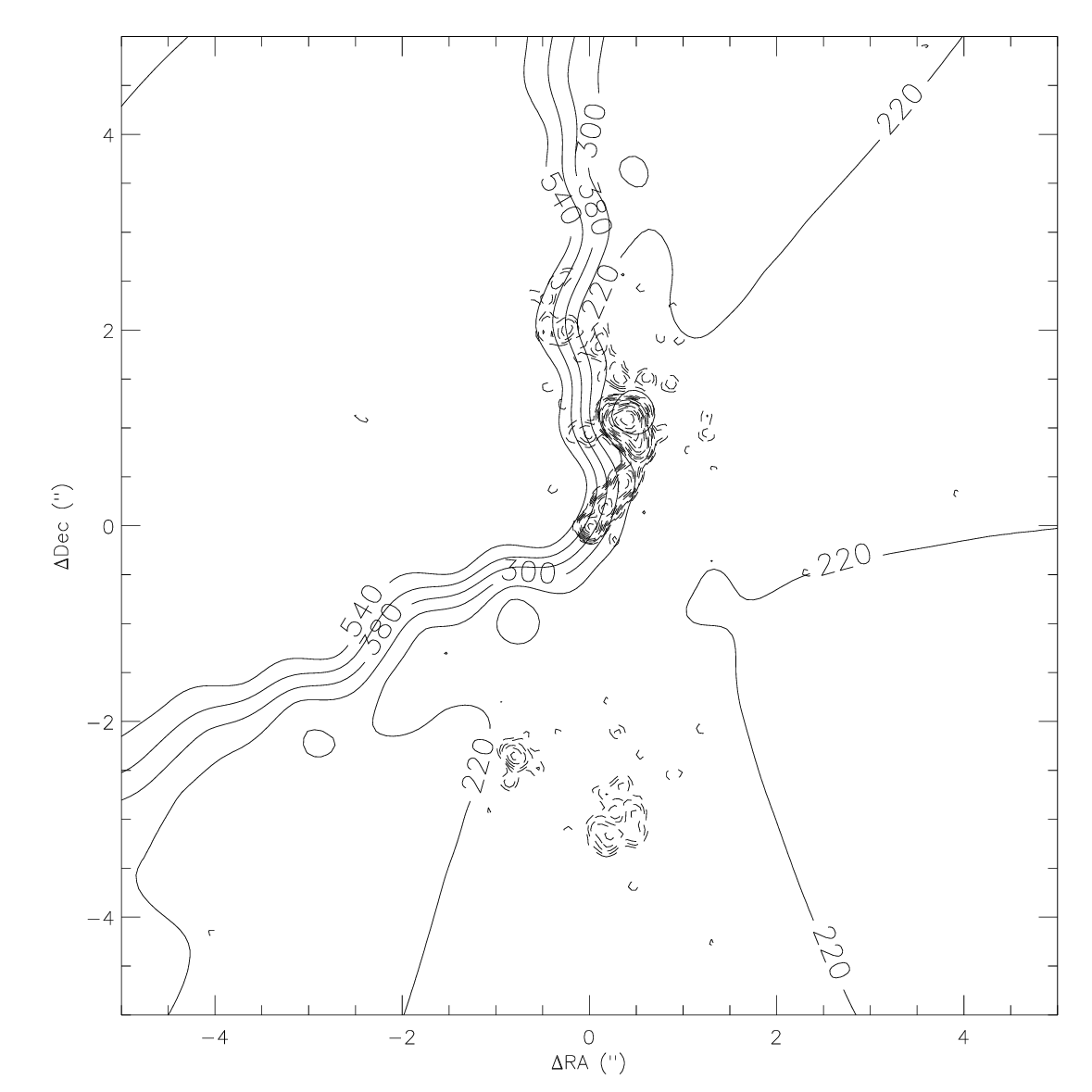}}
\caption{The predicted 21 cm maps for B2 0902+34. 
The 1.65 GHz data of \citet{Cari95} is plotted as the dashed contour as a likely source distribution.
\textit{Top Left} Column density where 
the log of N(HI) in cm$^{-2}$ is given. \textit{Top Right} 
Velocity dispersion in km s$^{-1}$. \textit{Bottom} 
Velocity shift in km s$^{-1}$.}
\label{duals}
\end{figure}

 Our model naturally reproduces the 21 cm absorption measurements without the need for a 
counterpart dip in the Lyman-$\alpha$ profile. By embedding the emission sites of 
Lyman-$\alpha$ in a larger neutral halo, the effects of scattering will not leave an absorption-like 
imprint on the Lyman-$\alpha$ profile, but they will do so on the 21 cm profile against the radio continuum emission. 
All existing 
21 cm data on B2 0902+34 are spatially unresolved. Figure \ref{duals} 
predicts the 21 cm maps that could be made using radio interferometry. The exact source distribution at 21 cm (323 MHz observed frame) is unknown, 
but we estimate it to be the same as the 1.65 GHz distribution of \citet{Cari95}. 
We assume the radio source 
plane to be the sky plane rotated onto the radio axis at an 
inclination of 45$\arcdeg$. We use the same 
model parameters as introduced in \S \ref{mod_eval}. 
We find a column density of 1.0$\times10^{21}$ to 1.0$\times10^{22}$cm$^{-2}$, 
a dispersion of 135 to 210 km s$^{-1}$, and a velocity 
shift from systemic of -190 to -275 km s$^{-1}$. These 
column densities would make this object a Damped Lyman-$\alpha$ system 
with a radius $>50$ kpc. The highest intensity radio emission 
resides is in a region of large gradients for the absorption properties, so slight changes 
in source distribution or rotation in the plane of the sky cause large 
changes in the integrated values. We give the ranges that misalignments of 
$\pm$1$\arcsec$ can create. The predicted line width is a factor of 2 above the 
measurement, but the rather unconstrained $\alpha$ parameter in the velocity field 
could likely remedy the mismatch as it takes a value closer to zero. 
Otherwise, the observations are bounded by these ranges, 
and the velocity shift tells us that 
our model self-consistently has both Lyman-$\alpha$ emission peaks blueward 
of both the 21 cm redshift and the systemic redshift. It is the velocity shift of 
190 km s$^{-1}$ for our fiducial astrometry that led us to the earlier assertion 
of $z_{systemic}=3.3990$. When building our resonant scattering model, we 
only let the 21 cm measurements guide our choice of systemic redshift. It is encouraging 
that our model designed to match the bimodal Lyman-$\alpha$ line profiles recovers so well the 
spatially unresolved 21 cm absorption signatures. 
\par Our most robust prediction comes from the southern radio lobe. 
Approximately 10\% of the radio signal originates from this region 
in a $\sim$1$\arcsec$ diameter aperture judging from the 1.65GHz data. 
Assuming the absorption strength is uniform over the halo, this aperture 
should have an absorption strength of 1.6mJy. Spatially 
resolved observations would need an unattainably long 
1100 hours of exposure with 4 dishes from the Very Long Baseline Array (VLBA) 
and the phased Very Large Array (VLA) 
to reach the same S/N of 6 as in the spatially unresolved absorption 
measurement \citep{Cody03} at a spectral 
resolution of 62.5kHz (restframe 58km s$^{-1}$) around the 323.053MHz observed frequency 
according to the European VLBI Network calculator\footnote[1]{http://www.evlbi.org/cgi-bin/EVNcalc}. 
Adding more VLBA dishes or other facilities, such as Greenbank or Arecibo, does not lower the 
exposure time as the increased baselines overresolve the southern lobe. However, 
the brighter northern jet could be measured at a S/N of 5 in 40 hours under the same 
configuration, and we are pursuing that observation. A spatially resolved 21cm measurement 
will give important confirming 
or falsifying evidence for a vast and massive infalling gas halo. 

\section{Implications: Infall Versus Outflow}
\label{in_or_out}
\par We are left with two serious, competing pictures: 
outflow from a 100 kpc HII region or infall of a 100 kpc HI region with embedded cones of HII.
Deciding between these mechanisms has impact beyond B2 0902+34. The bimodal line profile we see in 
B2 0902+34 is fairly common in HzRG halos. \Citet{vanO97} found 11 of 18 
such systems to carry the feature in longslit spectroscopy, although they all have not been studied for the 
confirming 2D spatial profiles as can be done with IFS. 
\par
\begin{figure}
\centering
\epsscale{1.0}
\includegraphics[angle=270,scale=0.6] {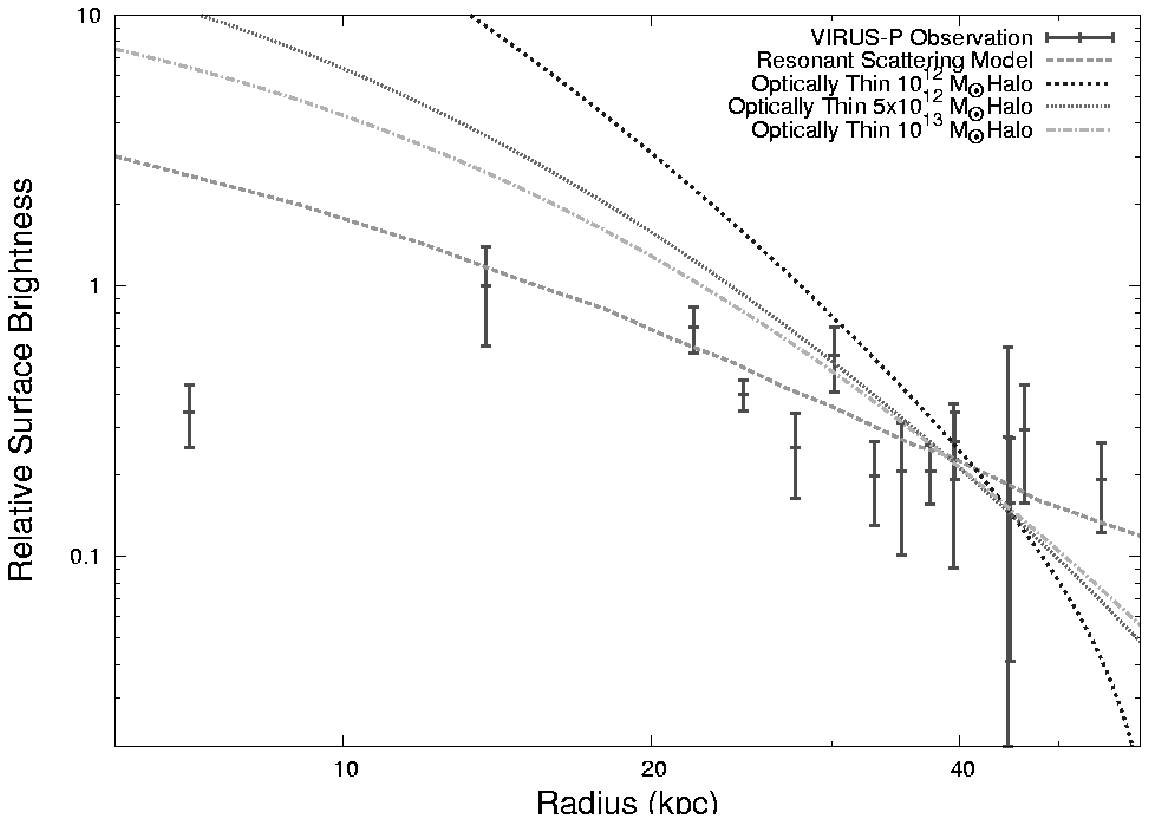}
\caption{Observed and simulated surface brightness profiles. All models have been 
scaled in intensity by least squares to the data. The optically thin 
integrals through NFW halos fail, though the resonant scattering model reproduces 
the observed, shallower profile.}
\label{fig5}
\end{figure}

 The outflow picture requires shells or clumps of HI at a large radius from the emission halo. This 
gas scatters a fraction of the line emission, and the blueshift in the absorption trough argues for 
outflow. The energy mechanism might be a starburst-driven superwind as discussed above. 
\citet{Wilm05} has applied such an explanation to LAB2 as has \citet{Reul07} for 
several HzRGs including B2 0902+34. If correct, these systems may be showing feedback in action with 
gas being heated and driven out. We see three reasons that this model is likely inappropriate to 
B2 0902+34. First, the 21 cm absorbing HI has no natural home in this picture. The radio emission, 
which extends to $\approx$40 kpc must lie behind this bit of HI, but the Lyman-$\alpha$ photons must be generated 
in front or must at least have many scatterings after the absorption 
as no spectra shows Lyman-$\alpha$ absorption at the 21 cm redshift and column 
density. \citet{Reul07} 
recognize this and predict a small, dense cloud of HI around the AGN. 
This too is unlikely, as the amount of spectrally flat radio core emission available 
for absorption in 
the data of \citet{Cari95} and the amount required by the 21cm measurement of 
\citet{Cody03} are modestly incompatible. Second, the surface brightness profile 
has the wrong slope for unscattered, extended emission. As shown in Figure \ref{fig5}, the 
observed surface brightness profile follows a power law. Taking an NFW profile for 
density and integrating the square of the density through the halo, assuming 
an optically thin case and highly extended photoionization, we find surface brightness 
profiles that are too steep for all reasonable halo masses. Our simulation of resonant 
scattering gives a nearly power law profile with an appropriate slope.
Third, the geometry of an absorbing 
cloud would have to be finely tuned to produce the observation. Based on our spectroscopy and 
that of \citet{Reul07}, the primary peak's center stays very constant at 5339.0 $\pm$ 2.0\AA\ 
in all positions. In the NE, there is no secondary emission peak so the absorption must 
be total and go quite blue. In the SW, the absorption would have to be weaker, narrower, and 
redder than in the NE to produce Figure 4 of \citet{Reul07}. The outflow picture does not 
produce a natural explanation for the spatial variation in the line profile.
\par Our result is surprising compared to the study of \citet{Nesv08} with data from 
three other HzRGs. 
The authors find extended emission in rest-frame optical lines and kinematics 
indicative of outflow. A possibly crucial difference between the datasets 
is that for B2 0902+34 the emission line regions extend far beyond the radio lobes, while 
the other three HzRGs are bounded by radio emission and significantly elongated along the 
radio axis. If we speculate that both our and their conclusions are true, the 
next questions becomes whether the same HzRGs show different dynamical signatures 
through different lines or whether different HzRGs show different dynamical signatures 
across all their lines. 
\par The infall picture has been suggested before for HzRGs \citep[e.g.][]{Hump07}. \citet{Vill07b} 
produced a model of optically thin biconal emission in Lyman-$\alpha$ to recover their observations 
of a peak emission position and peak velocity offset position. Their models, however, 
required postulation of a dense core that blocks emission from a rear cone 
that would have upset their trends, so their model is inappropriate to a bimodal 
line profile. They also could not produce lines at the width of their observations. 
We agree with the infall picture but use an optically thick regime of ionization cones embedded in one large, 
continuous neutral halo to explain all the available data on B2 0902+34 as we discussed above. If correct, this method 
has the capability to measure distributed hydrogen mass which we predict to be of the order 
$10^{12}M_{\odot}$. This is 5 orders of magnitude higher than the warm and cool hydrogen populations estimates by 
\citet{DeBr03} for the HzRG J2330+3927 from optically 
thin assumptions and Lyman-$\alpha$ line profile absorption specifically, 
and similarly high for the typical mass budget of HzRGs \citep{Mile08}. The neutral gas 
mass is dominant by a factor of $\ge$17 compared to the 
estimate of stellar mass in B2 0902+34 of $6\times10^{10}M_{\odot}$ \citep{Seym07}. We note that in the work of 
\citet{Seym07}, B2 0902+34 has an usually low stellar mass compared to other radio galaxies at a similar epoch. Combined with our work here, 
this low stellar mass may indicate HzRGs exist in a fairly heterogeneous range of galaxy evolution, with B2 0902+34 at an earlier stage of evolution 
than the average HzRG. The unusually early evolutionary state and large HI halo for B2 0902+34 may also 
be why only B2 0902+34 so far has shown 21 cm absorption.  
Our model also puts the system in a radically different evolutionary state from that predicted by outflow. Instead, we have 
infalling gas that is still building up what will become the galaxy's luminous mass. Star and 
dust formation have not yet built up to a dominant level to affect the Lyman-$\alpha$ line. The central AGN 
has not yet blown out or heated a significant fraction of its environment in an act of feedback. 
The model mass would place B2 0902+34 
as one of the larger halos that should collapse at the observed redshift. This picture does not 
deny HzRGs the role of feedback agents, but does show that radio emission can be output
in a protogalactic phase as well and that AGN buildup may predate significant star formation. 
Further study on B2 0902+34 may be important in deciding the origin of supermassive black holes 
\citep{Djor08}, 
and their coevolution with stellar bulges. Two likely paths of early black hole formation are as 
end biproducts of the earliest stars \citep{Mada01,Rico04} or as gas that avoided molecular cooling 
and fragmentation and collapsed 
into a black hole without associated star formation \citep{Silk98,Loeb94,Eise95,Brom03,Kous04}. 
In the latter case, a phase of galaxy evolution may exist where the stellar bulge and black hole 
masses are not tightly related. Although we do not have an estimate of B2 0902+34's black hole mass, the large 
radio continuum luminosity suggests an already large black hole. 
B2 0902+34, although we do not observe any associated cluster LAEs, may yet grow into a rich cluster as the normal 
cluster members begin their own star formation, possibly triggered by B2 0902+34 itself as 
described in \citet{Rawl04}.
\par While we advocate a resonantly scattering, infalling model to explain the extent of emission 
and the otherwise conflicting Lyman-$\alpha$ and 21cm absorption data, our model does 
require a minimum of $10^{12}M_{\odot}$ in HI subject to the modeled constraint of a maximum 
infall velocity of $v_{bulk}=v_{amp}\times v_{vir}\times (r/r_{vir})^\alpha$ with 
here $v_{amp}\le1.5$. 
Having this much HI inside a galaxy's 
virial radius conflicts with the classic galaxy formation scenario 
\citep{Binn77,Rees77,Silk77}. In that picture, 
a shock near the virial radius should ionize incoming gas and heat it to $\approx10^6K$ 
where it stays in quasi-hydrostatic equilibrium. Especially in the most massive halos, 
the low density gas is expected to have a cooling time longer than the galaxy age. This scenario 
has been adjusted in recent years with rapidly cooling accretion flows showing up in 
semi-analytic simulations \citep{Crot06} and cold accretion in 
smoothed particle hydrodynamics (SPH) simulations where a virial shock 
is mass conditionally unstable \citep{Birn03} and filaments provide many paths where cold gas ($<10^5K$) 
can flow to the forming galaxy's center \citep{Fard01,Kere05,Deke06,Deke08,Kere08,Ocvi08}. 
The ability of cooling and cold flows 
to exist have a redshift and halo mass dependence. \citet{Crot06} give 
$M_{halo}<2-3\times10^{11} M_{\odot}$ for z=0-6 and \citet{Kere05} give 
$M_{halo}<10^{11.4} M_{\odot}$ at z=0 with some redshift evolution. In either case, 
our advocacy for a $M_{halo}>5\times10^{12} M_{\odot}$ at z=3.4 appears to be in conflict 
with theory. Instead, the growth mechanism of such a massive halo is expected to be mergers. 
Future radiative transfer simulations of Lyman-$\alpha$ in more realistic, filament 
conditions may alleviate this mismatch in mass scale, although since the velocity fields 
in \citet{Kere05} are of the same order as those we modeled, an easy solution is not obvious. 
The potential conflict with theory 
at such a large mass makes B2 0902+34 a compelling target for further observations and simulations.  
\section{Conclusions}
\label{Conc_sec}
\par We have made new spatially resolved IFS observations of Lyman-$\alpha$ in B2 0902+34. We find 
a spatially resolved region of weaker and bluer emission superimposed on a symmetric 
halo of primary emission. We have shown that all properties of B2 0902+34 can best be fit with 
a resonant scattering infall model with a large, mostly neutral, hydrogen mass. The 
sign of the velocity field is relatively robust and model independent. Signal seen 
through higher optical depths becomes increasingly bluer indicative of infall and 
resonant scattering. We match our data with a model dependent HI mass 
of $\ge10^{12}M_{\odot}$ 
which is $\ge$17$\times$ larger than the stellar mass. 
For three reasons an outflowing HI scenario cannot fit our observations: 
1) the strength of the observed 21cm absorption and the Lyman-$\alpha$ profile are incompatible 
with a single HI population if optically thin and the velocity offsets between the 21cm signal and the 
higher optical depth Lyman-$\alpha$ signal carry the wrong sign for outflow if optically thick, 
2) the surface brightness 
profile, with the assumption that the gas follows the dark matter, would be steeper than 
observed if optically thin, and 3) the bimodal Lyman-$\alpha$ line profile exists 
only in a small, non-central region of the galaxy which is incompatible with 
an outflowing shell geometry. 
The 21 cm HI absorption has been a powerful confirmation of the kinematics 
through its redshift offset from the Lyman-$\alpha$ halo emission, and we believe observational 
progress in understanding this class of object will benefit from further deep 21 cm 
measurements on other 
HzRGs. This new scenario of a massive, infalling HI halo with resonant scattering places B2 0902+34 
in a much earlier evolutionary state than the previously formulated starburst superwind picture, and 
we classify B2 0902+34 as a protogiant elliptical galaxy in an early collapse phase before 
significant feedback has disrupted star formation and AGN fueling.  
Observations of other HzRGs with IFS will be important 
in correctly classifying systems between formation and feedback modes and allow detailed 
observational tests of these two processes. We are observing a large sample of HzRGs 
with VIRUS-P in order to undertake a similar analysis on other systems.
\par Further types of data on this system will allow confirmation of our infalling, resonant 
scattering model. Polarization 
measurements of Lyman-$\alpha$ could confirm the scattering process \citep{Lee98}. 
The bright regions of HzRGs have been found to have modest levels of polarization 
($\approx$10\%, \citet{Vern01}) while scattering simulations in similar situations 
to our model \citep{Dijk08} 
predict higher ($\approx$40\%) values. A measurement between HzRGs with and without emission 
line halos may demonstrate this split. 
The key, potentially falsifying observation 
for this resonant scattering, infalling model of B2 0902+34 will be radio interferometry spectral 
imaging to spatially resolve the distribution of HI gas through its 21 cm absorption. 
\citet{Cari95} discussed the utility of such a measurement, and we have now given 
an observationally motivated model that predicts spatially extended HI.
\acknowledgments

This material is based upon work supported under a National Science 
Foundation Graduate Research Fellowship. This work is supported 
in part by Texas Advanced Research Project Grant number 003658-0005-2006. 
The simulations used in this work were carried out at the Texas Advanced Computing Center (TACC). 
The authors would like to thank Karl Gebhardt for discussion and code, Guillermo Blanc for assistance 
with the observations reported here, Chris Carilli and Carlos De Breuck 
for sharing their radio imaging, and an anonymous referee for
improvements in presentation and substance. 

{\it Facilities:} \facility{Smith (VIRUS-P)}.


\appendix
\section{Tests of the resonant scattering code}
\par
\begin{figure}
\centering
\epsscale{1.0}
\subfigure{\includegraphics[angle=270,scale=0.3] {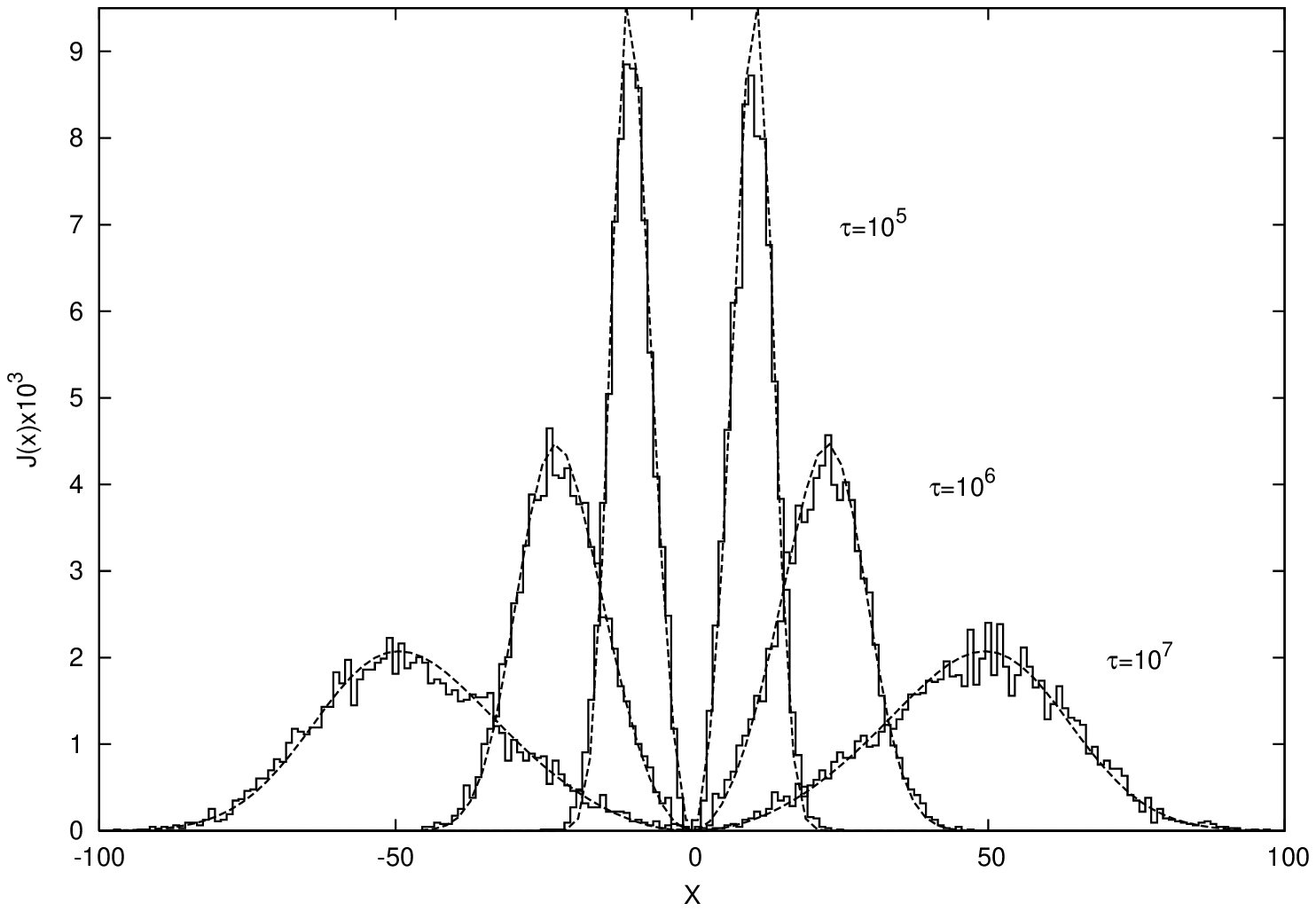}}
\subfigure{\includegraphics[angle=270,scale=0.3] {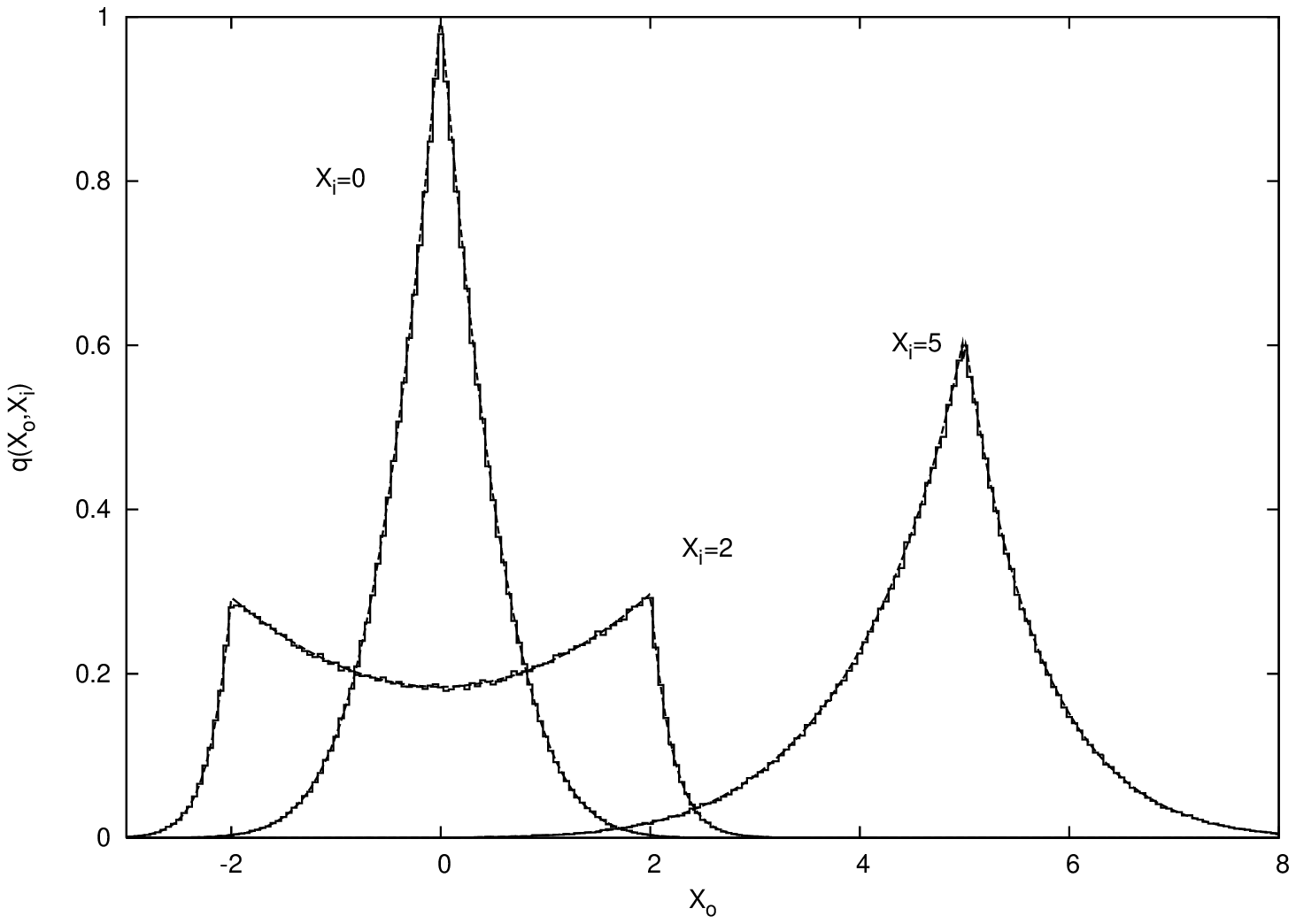}}
\\
\subfigure{\includegraphics[angle=270,scale=0.3] {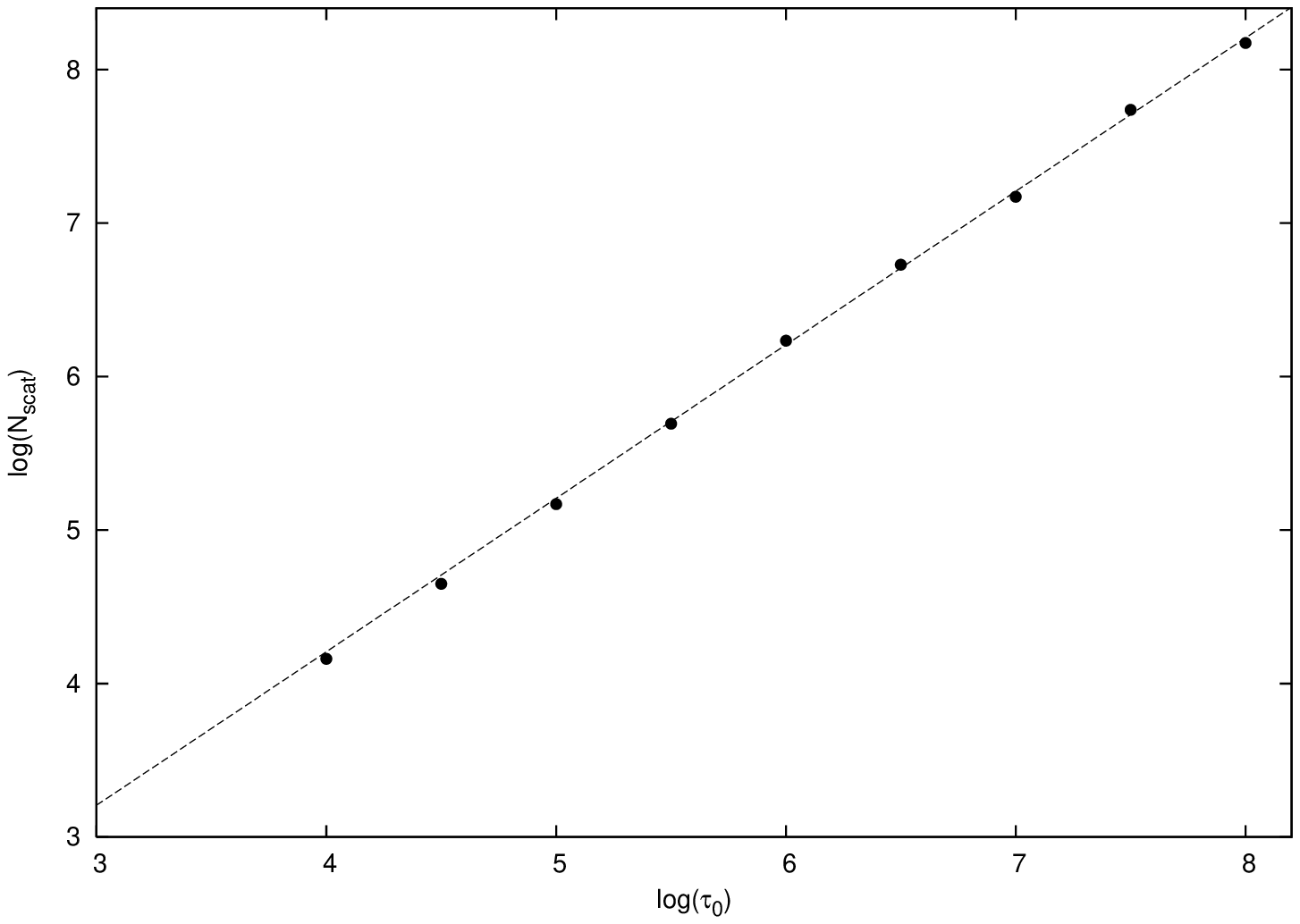}}
\subfigure{\includegraphics[angle=270,scale=0.3] {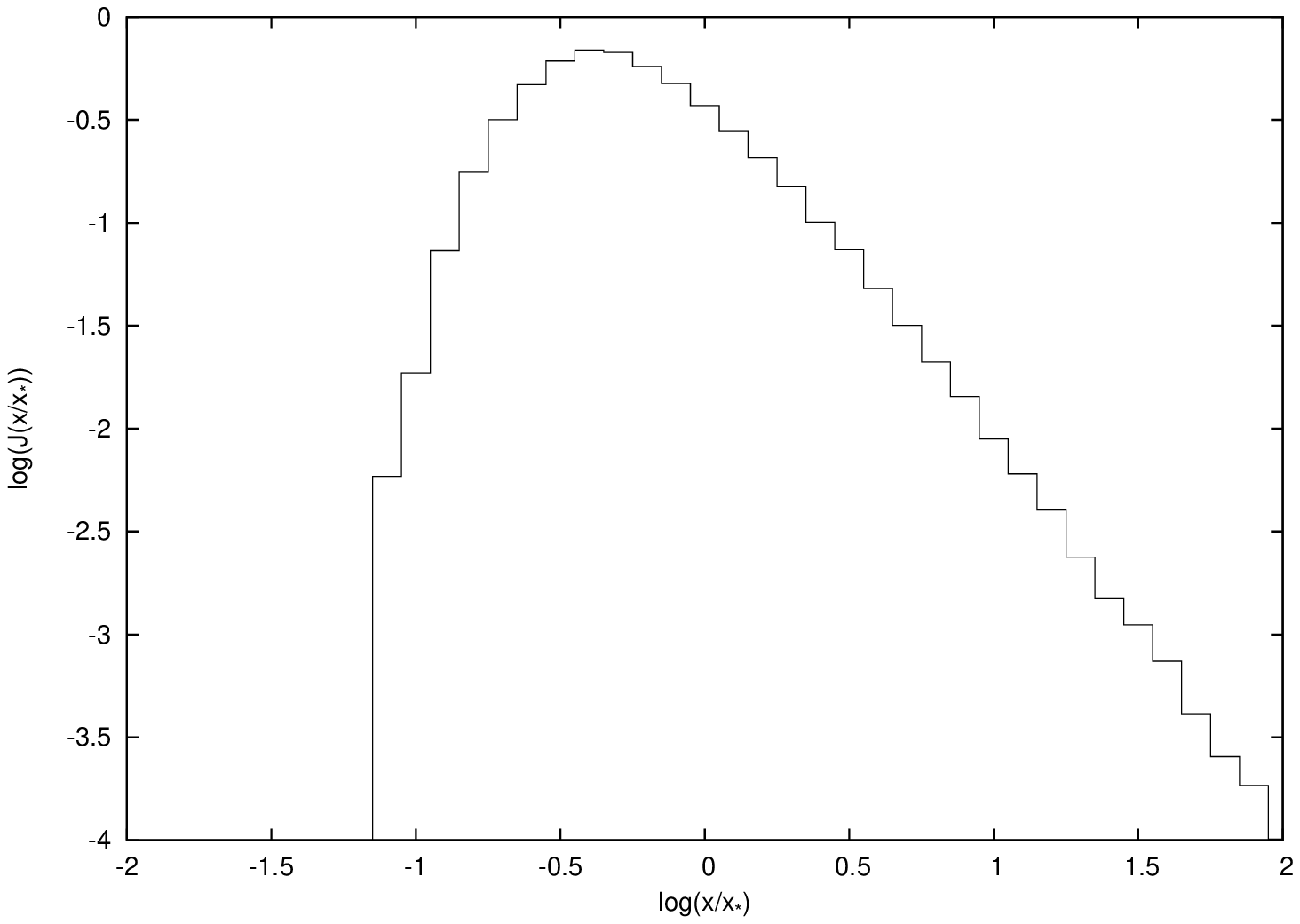}}
\caption{The four tests of our 3D Monte Carlo resonant scatter code. 
\textit{Top left} The emergent Lyman-$\alpha$ spectra from uniform density, static halos at 
T=10K with various line center edge-to-center optical depths. The histograms are 
from our simulations. The lines are from function 9 in \citet{Dijk06}. 
 \textit{Top right} The redistribution function 
for Lyman-$\alpha$ scattering with a gas temperature of T=10$^4$K. The 
histogram is our code output while the curves are from theory in \citet{Humm62,Lee74}.  
\textit{Bottom left} The number of scatterings necessary for a Lyman-$\alpha$ 
photon to escape a neutral, static slab of HI. The points are from our 
simulations. The line is from theory \citep{Harr73}. \textit{Bottom right} 
The emergent Lyman-$\alpha$ spectrum from an infinite, uniform density HI halo 
under Hubble expansion at z=10 with T=10K.
}
\label{fig_restest}
\end{figure}

 To provide compatibility between our work and previous authors', we run the 
same set of four tests on our 3D Monte Carlo resonant scattering code as 
\citet{Dijk06}. The first test simulates the emergent spectra for centrally 
created Lyman-$\alpha$ photons in a uniform density spherically symmetric static halo of HI 
with various line center edge-to-center optical depths at T=10K. We disabled the 
acceleration scheme for tests 1-3. 
The second test calculates the redistribution function for single 
scattering events, which is the probability distribution function for an output 
frequency given an input frequency and gas temperature. The third test finds the mean 
number of scatterings prior to escape for a slab of gas of various line center 
optical depths. The weighting scheme of \citet{Aver68} was used for test 3 where 
the number of accumulated scatterings and the probability of escape from each 
scattering site is retained as opposed to solely keeping the properties of 
each Monte Carlo run's escaping photon.  
The fourth test gives the emergent spectrum from an infinitely 
large object under Hubble expansion as originally simulated 
in \citet{Loeb99}. More details and references are given in 
\citet{Dijk06} Appendix B. We give the four tests in Figure \ref{fig_restest}.


\end{document}